\documentclass[journal]{IEEEtranTIE}

\usepackage{subcaption}

\usepackage{bm}
\usepackage{diagbox}

\newcommand{\identityMatrix}{\pmb{I}}
\newcommand{\zeroMatrix}{\pmb{0}}
\newcommand{\integerNumbers}{\mathbb{Z}}
\newcommand{\realNumbers}{\mathbb{R}}

\newcommand{\module}[1]{\left\|{#1}\right\|}
\newcommand{\abs}[1]{|{#1}|}
\newcommand{\minEigenvalue}[1]{\lambda_{\textrm{min}}({#1})}
\newcommand{\maxEigenvalue}[1]{\lambda_{\textrm{max}}({#1})}

\newcommand{\referenceVoltage}{v_r}
\newcommand{\referenceVoltageDerivative}{\dot{v}_r}
\newcommand{\outputVoltage}{v_o}
\newcommand{\referenceVoltageSecondDerivative}{\ddot{v}_r}
\newcommand{\referenceVoltageThirdDerivative}{\dddot{v}_r}
\newcommand{\outputVoltageSecondDerivative}{\ddot{v}_o}
\newcommand{\totalDisturbance}{F}
\newcommand{\errorDomainTotalDisturbance}{F^*}
\newcommand{\errorDomainTotalDisturbanceDerivative}{\dot{F}^*}
\newcommand{\errorDomainTotalDisturbanceObservationError}{\tilde{F}^*}
\newcommand{\errorDomainTotalDisturbanceEstimate}{\hat{F}^*}
\newcommand{\extendedState}{\bm{z}}
\newcommand{\extendedStateElement}[1]{{z}_{#1}}
\newcommand{\extendedStateEstimateElement}[1]{\hat{{z}}_{#1}}

\newcommand{\extendedStateAggregatedObservationError}{\tilde{\bm{\zeta}}}
\newcommand{\extendedStateAggregatedObservationErrorDerivative}{\dot{\tilde{\bm{\zeta}}}}
\newcommand{\extendedStateAggregatedStateMatrix}{\bm{H}_\zeta}
\newcommand{\aggregatedDisturbanceInput}{\bm{\delta}}
\newcommand{\aggregatedNoiseVector}{\bm{\gamma}}
\newcommand{\extendedStateObservationErrorCascadeLevel}[1]{\tilde{\bm{z}}_{#1}}
\newcommand{\extendedStateObservationErrorCascadeLevelElement}[1]{\tilde{{z}}_{#1}}
\newcommand{\extendedStateObservationErrorCascadeLevelDerivative}[1]{\dot{\tilde{\bm{z}}}_{#1}}
\newcommand{\extendedStateEstimate}{\hat{\bm{z}}}
\newcommand{\extendedStateDerivative}{\dot{\bm{z}}}
\newcommand{\controlError}{e}
\newcommand{\controlErrorDerivative}{\dot{e}}
\newcommand{\controlErrorSecondDerivative}{\ddot{e}}
\newcommand{\cgpEstimate}{\hat{b}}
\newcommand{\controlSignal}{\mu}
\newcommand{\stabilizingController}{\mu_0}
\newcommand{\stateMatrix}{\pmb{A}}
\newcommand{\inputVector}{\pmb{d}}
\newcommand{\disturbanceInputVector}{\pmb{b}}
\newcommand{\outputVector}{\pmb{c}}
\newcommand{\stateTransformationMatrix}[1]{\pmb{\Lambda}_{#1}}

\newcommand{\transformedObservationErrorStateMatrix}{{\pmb{H}_\chi}}

\newcommand{\observationErrorStateMatrix}[1]{\pmb{H}_{\chi}}

\newcommand{\observerStageState}[1]{\pmb{\xi}_{#1}}
\newcommand{\observerStageStateElement}[1]{{\xi}_{#1}}

\newcommand{\observerStageStateDerivative}[1]{\dot{\pmb{\xi}}_{#1}}
\newcommand{\observerGainVector}[1]{\pmb{l}_{#1}}

\newcommand{\cascadeLevel}{p}
\newcommand{\observerStageBandwidth}[1]{\omega_{o#1}}
\newcommand{\observerBandwidthMultiplier}{\alpha}

\newcommand{\stabilizingControllerProportionalGain}{k_p}
\newcommand{\stabilizingControllerDerivativeGain}{k_d}

\newcommand{\observationErrorUltimateBound}{\delta_{\tilde{z}}}
\newcommand{\controlErrorUltimateBound}{\delta_{e}}
\newcommand{\systemOutput}{y}
\newcommand{\measurementNoise}{n}

\newcommand{\modifiedExtendedStateAggregatedObservationError}{\bm{\chi}}
\newcommand{\modifiedExtendedStateAggregatedObservationErrorDerivative}{\dot{\bm{\chi}}}
\newcommand{\observerLevelTransformationMatrix}[1]{\pmb{L}_{#1}}

\newcommand{\observationErrorLyapunovFunction}{V_\chi}
\newcommand{\observationErrorLyapunovFunctionDerivative}{\dot{V}_\chi}
\newcommand{\observationErrorLyapunovEquationSolution}{\pmb{P}_\chi}
\newcommand{\observationErrorMajorizationConstant}{\nu_\chi}

\newcommand{\kappaFunction}{\mathcal{K}}
\newcommand{\controlErrorVector}{\pmb{\epsilon}}
\newcommand{\controlErrorVectorDerivative}{\dot{\pmb{\epsilon}}}
\newcommand{\controlErrorGainMatrix}{\pmb{K}}
\newcommand{\stabilizingControllerParameter}{k}
\newcommand{\transformedControlError}{\pmb{\varepsilon}}
\newcommand{\transformedControlErrorDerivative}{\dot{\pmb{\varepsilon}}}
\newcommand{\controlErrorPerturbationMatrix}{\pmb{Z}}
\newcommand{\controlErrorPerturbationMatrixMaxValue}{m_Z}
\newcommand{\transformedControlErrorStateMatrix}{\pmb{H}_\varepsilon}
\newcommand{\transformedControlErrorLyapunovEquationSolution}{\pmb{P}_\varepsilon}
\newcommand{\transformedControlErrorLyapunovFunction}{V_\varepsilon}
\newcommand{\transformedControlErrorLyapunovFunctionDerivative}{\dot{V}_\varepsilon}
\newcommand{\transformedControlErrorMajorizationConstant}{\nu_\varepsilon}
\newcommand{\measurementNoiseInputVector}{\pmb{\kappa}}
\newcommand{\ls}{{\textrm{ls}_\infty}}

\newcommand{\eigenvalue}[1]{\lambda_{#1}}
\newcommand{\extendedStateDomain}{\mathcal{D}_z}

\usepackage{graphicx}
\usepackage{cite}
\usepackage{picinpar}
\usepackage{amsmath}
\usepackage{url}
\usepackage{flushend}
\usepackage{colortbl}
\usepackage{soul}
\usepackage{multirow}
\usepackage{pifont}
\usepackage{color}
\usepackage{alltt}
\usepackage[hidelinks]{hyperref}
\usepackage{enumerate}
\usepackage{siunitx}
\usepackage{breakurl}
\usepackage{epstopdf}
\usepackage{pbox}
\usepackage{amsfonts}

\usepackage{mathtools}
\usepackage{MnSymbol}
\usepackage{graphicx}
\usepackage{lipsum}
\usepackage{stackrel}

\newtheorem{remark}{Remark}
\newtheorem{theorem}{Theorem}
\newtheorem{assumption}{Assumption}

\usepackage[table]{xcolor}

\usepackage[pscoord]{eso-pic}% The zero point of the coordinate systemis the lower left corner of the page (the default).

\newcommand{\placetextbox}[3]{% \placetextbox{<horizontal pos>}{<vertical pos>}{<stuff>}
  \setbox0=\hbox{#3}% Put <stuff> in a box
  \AddToShipoutPictureFG*{% Add <stuff> to current page foreground
    \put(\LenToUnit{#1\paperwidth},\LenToUnit{#2\paperheight}){\vtop{{\null}\makebox[0pt][c]{#3}}}%
  }%
}%

\newcommand{\addheaders}{\placetextbox{0.5}{1}{\scriptsize \color{gray}This is the author’s version of an article that has been accepted for publication in this journal. Changes will be made to this version by the publisher prior to final publication.}%
\placetextbox{0.5}{0.99}{\scriptsize \color{gray}The official version of record is available at \color{customblue}\url{http://dx.doi.org/10.1109/TIE.2021.3055187}}%

\placetextbox{0.5}{0.05}{\scriptsize \color{gray}\copyright2021 IEEE. Personal use of this material is permitted. Permission from IEEE must be obtained for all other uses, in any current or future media, including reprinting/republishing this material}%
\placetextbox{0.5}{0.04}{\scriptsize \color{gray}
for advertising or promotional purposes, creating new collective works, for resale or redistribution to servers or lists, or reuse of any copyrighted component of this work in other works.}}%}

\definecolor{customblue}{rgb}{0.447,0.639,0.949}

\begin{document}

\addheaders
%  \title{Active Disturbance Rejection Control\\with Sensor Noise Suppressing Observer\\for DC-DC Buck Power Converters}
%  \title{\textcolor{black}{Active Disturbance Rejection Control Design with Sensor Noise Effect Suppression in Application to DC-DC Buck Power Converter}}
 \title{{Active Disturbance Rejection Control Design with Suppression of Sensor Noise Effects in Application to DC-DC Buck Power Converter}}

\author{Krzysztof~Łakomy,
        Rafal~Madonski,
        Bin~Dai,
        Jun~Yang,~\IEEEmembership{Senior Member,~IEEE},
        Piotr~Kicki,
        Maral~Ansari,~\IEEEmembership{Student Member,~IEEE},
        Shihua~Li,~\IEEEmembership{Fellow,~IEEE}

\thanks{
Manuscript received Month xx, 2xxx; revised Month xx, xxxx; accepted Month x, xxxx. %The article was created thanks to participation in program PROM of the Polish National Agency for Academic Exchange. The program is co-financed from the European Social Fund within  the Operational Program Knowledge Education Development, non-competitive project entitled “International scholarship exchange of PhD students and academic staff” executed under the Activity 3.3 specified in the application for funding of project No. POWR.03.03.00-00-PN13/18. The work has been also supported by the Fundamental Research Funds for the Central Universities project no.~21620335.

K.~Łakomy is with the Poznań University of Technology, Poznań, 60-965, Poland (e-mail: krzysztof.lakomy92@gmail.com).

R.~Madonski (corresponding author) is with the Energy Electricity Research Center, International Energy College, Jinan University, Zhuhai, 519070, P. R. China (e-mail: rafal.madonski@jnu.edu.cn).

B.~Dai, J.~Yang and S.~Li are with the School of Automation, Southeast University, Key Laboratory of Measurement and Control of CSE, Ministry of Education, Nanjing, 210096, P. R. China (e-mail: \{bin\_1994/j.yang84/lsh\}@seu.edu.cn).

P.~Kicki is with the Institute of Robotics and Machine Intelligence, Poznań University of Technology, Poznań, 60-965, Poland (e-mail: piotr.kicki@put.poznan.pl).

M.~Ansari is with the Faculty of Engineering and Information Technology, University of Technology Sydney, Sydney NSW, 2007, Australia (e-mail: maral.ansari@student.uts.edu.au).
}
}

\markboth{IEEE TRANSACTIONS ON INDUSTRIAL ELECTRONICS}%
{}

\maketitle

\begin{abstract}
    The performance of active disturbance rejection control (ADRC) algorithms can be limited in practice by high-frequency measurement noise. In this work, this problem is addressed by transforming the high-gain extended state observer (ESO), which is the inherent element of ADRC, into a new cascade observer structure. Set of experiments, performed on a DC-DC buck power converter system, show that the new cascade ESO design, compared to the conventional approach, effectively suppresses the detrimental effect of sensor noise over-amplification while increasing the estimation/control performance. The proposed design is also analyzed with a low-pass filter at the converter output, which is a common technique for reducing measurement noise in industrial applications.
\end{abstract}

%%%%%%%%%%%%%%%%%%%%%%%%%%%%%%%%%%%%%%%%%
%%%%%%%%%%%%%%%%%%%%%%%%%%%%%%%%%%%%%%%%%
%%%%%%%%%%%%%%%%%%%%%%%%%%%%%%%%%%%%%%%%%
\section{Introduction}

\IEEEPARstart{R}{enewable} energy sources, like fuel and photovoltaic cells, are rapidly evolving technologies for DC voltage generation, which results in proliferation of DC–DC buck converters in power applications. Practically appealing results on buck converter control using the idea of active disturbance rejection control (ADRC) were recently reported in~\cite{NaszeCEP,WuHanTIE,TransCircuits2}.
% Interestingly, some motor control companies attracted by the ADRC-based solutions (e.g. Texas Instruments), have embedded this approach in their selected commercial products~\cite{Instaspin}.
% ESOsmcDCDC,TransCircuits3,TransCircuits4
% \cite{TransCircuits2,TransCircuits3,TransCircuits4,SiraDCDC,FTDOBdcdc,ESOsmcDCDC}
% references (IEEE Transactions on Circuits and Systems):
% \begin{itemize}
%     \item \cite{TransCircuits1}: Offset-Free Nonlinear {MPC} for Mismatched Disturbance Attenuation With Application to a Static Var Compensator
%     \item \cite{TransCircuits2}: Optimized Active Disturbance Rejection Control for DC-DC Buck Converters With Uncertainties Using a Reduced-Order GPI Observer
%     \item \cite{TransCircuits3}: GPIO-based robust control of nonlinear uncertain systems under time-varying disturbance with application to DC–DC converter
%     \item \cite{TransCircuits4}: Design and Qualitative Robustness Analysis of an DOBC Approach for DC-DC Buck Converters With Unmatched Circuit Parameter Perturbations
%     \item \cite{TransCircuits5}: Continuous output feedback TSM control for uncertain systems with a DC-AC inverter example
%     \item \cite{TransCircuits6}: Robust Voltage Regulation of a DC–AC Inverter With Load Variations via a HDOBC Approach
% \end{itemize}
The key element in any ADRC scheme is the extended state observer (ESO\cite{Han-fromPID}), responsible for estimating the system state vector and reconstructing the overall disturbance (also referred to as \textit{total disturbance}) affecting the controlled variable~\cite{SenChen}.

However, since the conventional form of ADRC uses a high-gain observer (HGO) structure to estimate selected signals, its capabilities are intrinsically limited by the presence and severity of high-frequency sensor noise, as discussed in~\cite{DOBCoverview,obs-survey2,KLESOarchit}.  The high gains of the observer cause the transfer of strongly amplified measurement noise into the control signal calculated upon the state vector of ESO. This may cause the decrease of control quality (e.g. when the noise-affected control signal hits the actuator saturation), higher energy consumption, and quicker wear of the equipment.
% \color{black}["The importance of the problem considered in this paper should be further addressed"].
\color{black}
The HGO-based ADRC design and tuning often come down to a forced compromise between speed/accuracy of signals reconstruction and sensitivity to noise~\cite{khalil2014}. Same compromise can be seen in the ADRC works for buck converters in which the measured system output (voltage) is oftentimes corrupted with high-frequency noise~\cite{Powernoise}. Several types of solutions were proposed to solve the problem of attenuating the effects of measurement noise in high-gain observers. They mainly address it by: employing nonlinear~\cite{Han-fromPID,KhalinNonlineaObs} or adaptive techniques~\cite{balotelli}, redesigning the local behavior by combining different types of observers~\cite{WenchaoESFSunLi}, using low-power structures~\cite{ESOMarconi,limiPowerhighga,KhalilCascade}, or modifying standard low-pass filters~\cite{OutputInjecFilt}.

% In this work, a novel error-domain cascade ESO structure is used.
% \color{black}["The novelty of the proposed method should be highlighted in the introduction part."]. \color{black}
% initial observation from/made in.

Motivated by the above problem, a new cascade ESO-based error-domain ADRC solution is presented. Following the general idea shown in~\cite{lakomy2020cascade}, we propose a virtual decomposition
% It is based on a virtual decomposition
of the total disturbance present in the DC-DC buck converter system, allowing to design a cascade structure of ESO, where each level of the observer cascade is responsible for handling a particular type and frequency range of estimated signal. The proposed topology enhances conventional state/disturbance estimation performance while avoiding over-amplification of the sensor noise. The user-defined number of cascade levels allows to customize the overall control system structure to meet certain disturbance rejection requirements. Although a multi-level cascade observer is proposed, a straightforward design and implementation methodology is given, together with intuitive tuning rules.
The novelty of this work includes an experimental validation of the proposed cascade ESO-based ADRC structure, a proof of the input-to-state stability of the closed-loop system, and additional insights about the sensor noise suppressing effects in frequency domain.
\textcolor{black}{The experimental study also addresses the impact of a low-pass filter implemented at the converter output, which is a popular approach for handling high-frequency sensor noise~\cite{khalil2016}.}
% \textcolor{black}{The last one includes an investigation of the relation between the proposed design and a low-pass filter, which is a popular tool for handling high-frequency noise~\cite{khalil2016}.}
% , and additionally commented on using frequency-domain properties.
% The closed-loop behavior of the proposed cascade ESO-based ADRC controller \color{black} designed for a DC-DC buck converter system was proved to be input-to-state stable, and followed with some comments addressing the controller behavior in the frequency domain, and validated with hardware experiments.
%conducted on a DC-DC buck converter laboratory testbed.
% , ...which are considered the novelty of this work.

% NOVELTY w.r.t. ISAT
% - error-based
% - closed-loop analysis
% - experiment
% - freq. analysis

% compared to our original/initial paper on CESO \cite{} , the novelty of this work is...

% In light of the ava

% In this paper, we have extended the introductory work on the cascade observer structure \cite{lakomy2020cascade} by providing its full analysis.

% Before the new observer-based control method is introduced, some preliminary information is given next, including the standard ADRC which is is revisited as it will be used later for comparison study.

\textit{Notation.} Within this article, we treat $\realNumbers$ as a set of real numbers, $\mathbb{R}_{+}=\{x\in\realNumbers:x>0\}$ as a set of positive real numbers, $\mathbb{R}_{\geq0}=\{x\in\realNumbers:x>0\}$ as a set of non-negative real numbers, $\integerNumbers$ as a set of integers, $\minEigenvalue{\pmb{A}} \ \textrm{and} \ \maxEigenvalue{\pmb{A}}$ are respectively the minimal and maximal eigenvalues of matrix $\pmb{A}$, while $\pmb{A}\succ0$ means that matrix $\pmb{A}$ is positive definite. Function $f(x):\realNumbers\rightarrow\realNumbers$ belongs to class $\kappaFunction$ when it is strictly increasing and $f(0)=0$. The expression  $\ls:=\limsup_{t\rightarrow\infty}$ is used for the sake of notation compactness.

%%%%%%%%%%%%%%%%%%%%%%%%%%%%%%%%%%%%%%%%%
%%%%%%%%%%%%%%%%%%%%%%%%%%%%%%%%%%%%%%%%%
%%%%%%%%%%%%%%%%%%%%%%%%%%%%%%%%%%%%%%%%%
\section{Preliminaries}
%%%%%%%%%%%%%%%%%%%%%%%%%%%%%%%%%%%%%%%%%
%%%%%%%%%%%%%%%%%%%%%%%%%%%%%%%%%%%%%%%%%
%%%%%%%%%%%%%%%%%%%%%%%%%%%%%%%%%%%%%%%%%

%%%%%%%%%%%%%%%%%%%%%%%%%%%%%%%%%%%%%%%%%
%%%%%%%%%%%%%%%%%%%%%%%%%%%%%%%%%%%%%%%%%
%%%%%%%%%%%%%%%%%%%%%%%%%%%%%%%%%%%%%%%%%
\subsection{Simplified plant model and control objective}
%%%%%%%%%%%%%%%%%%%%%%%%%%%%%%%%%%%%%%%%%
%%%%%%%%%%%%%%%%%%%%%%%%%%%%%%%%%%%%%%%%%
%%%%%%%%%%%%%%%%%%%%%%%%%%%%%%%%%%%%%%%%%

\begin{figure}
	\centering
		\includegraphics[width=0.455\textwidth]{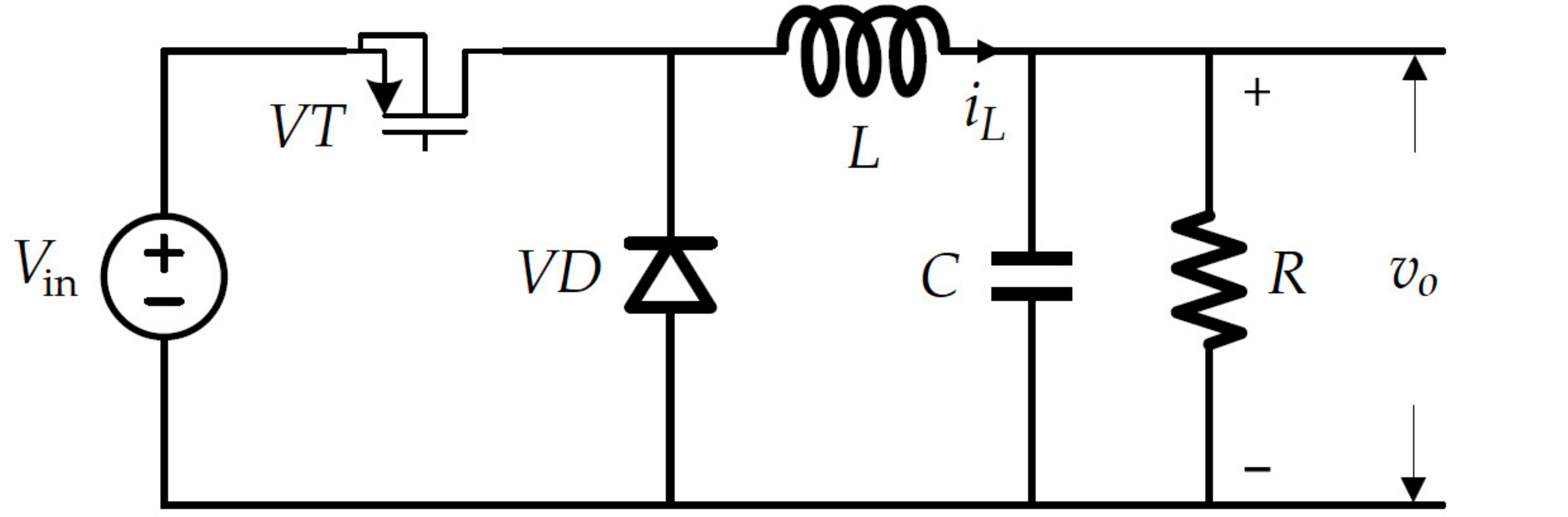}
	\caption{Semiconductor realization of the considered DC-DC buck power converter, with diode $VD$ and control switch $VT$.}
	\label{fig:DCDC}
\end{figure}

% ~\cite{SiraBookDCDC}
Following~\cite{TransCircuits2}, an \textit{average} dynamic model of a DC-DC buck converter, depicted in Fig.~\ref{fig:DCDC}, can be written as
% 	\begin{equation}
% 		\text{ON case~:}~\begin{cases}
% 		C\frac{dv_o(t)}{dt} = i_L(t)-\frac{1}{R}v_o(t),\\
% 		L\frac{di_L(t)}{dt} = V_{\text{in}}-v_o(t),
% 			\end{cases}
% 	\end{equation}
% 	\begin{equation}
% 		\text{OFF case~:}~\begin{cases}
% 		C\frac{dv_o(t)}{dt} = i_L(t)-\frac{1}{R}v_o(t),\\
% 		L\frac{di_L(t)}{dt} = -v_o(t),
% 			\end{cases}
% 	\end{equation}
% where the above switching structure can be also expressed with a following \textit{average} model, as explained in~\cite{SiraBookDCDC}:
	\begin{equation}
		\begin{cases}
		\frac{dv_o(t)}{dt} = \frac{1}{C}i_L(t) - \frac{1}{CR}v_o(t),\\
		\frac{di_L(t)}{dt} = \frac{V_{\text{in}}}{L}\left[\mu(t)+ d(t)\right] - \frac{1}{L}v_o(t), \\
		~y_o(t) = v_o(t) + n(t),
		\label{eq:DCDCmodel}
			\end{cases}
	\end{equation}
where $\mu \in [0,1]$ is the duty ratio, $y_o$[V] is the measured system output that consists of the average capacitor voltage $v_o$[V] and the sensor noise $n$[V], $i_L$[A] is the average inductor current, $R$[$\Omega$] is the load resistance of the circuit, $L$[H] is the filter inductance, $C$[F] is the filter capacitance, $V_{\text{in}}$[V] is the input voltage source, and $d(t)$ represents the unknown (possibly time-varying and nonlinear) external disturbance.

% and parametric discrepancies affecting the actual plant.

The considered control objective is to force $v_o(t)$ to follow a reference capacitor output voltage trajectory $v_r(t)$[V] by manipulating $\mu(t)$ with following assumptions applying.

\begin{assumption}
    \label{ass:CurrentVoltage}
    Following the limitations resulting from the physical properties of the considered electronic circuit, we may assume that the  values of voltage and current are bounded, and belong to some compact set such that $\sup_{t\geq0}|i_L(t)|<r_{i_L}$ and $\sup_{t\geq0}|v_o(t)|<r_{v_o}$ for $r_{i_L},r_{v_o}>0$.
\end{assumption}

\begin{assumption}
    \label{ass:Output}
    Output voltage $v_o(t)$ is the only measurable signal and is additionally corrupted by bounded, high-frequency measurement noise $\sup_{t\geq0}|n(t)|<r_n$  for $r_n>0$.
\end{assumption}
% \begin{assumption}
%     \label{ass:ExtDistBounded}
%     The unknown external disturbance signal $\sup_{t\geq0}|d(t)|<r_d$ is bounded and has bounded first time-derivative $\sup_{t\geq0}|\dot{d}(t)|<r_{\dot{d}}$.
% \end{assumption}

\begin{assumption}
    \label{ass:ExtDistBounded}
    \cite{huang2014} The unknown external disturbance~$d(t)$ may have a countable number of first-class discontinuity points\footnote{Function $f(x):\realNumbers\rightarrow\realNumbers$ has first-class discontinuity at point $\bar{x}$ if for $f^+:=\lim_{x\rightarrow\bar{x}^+}f(x)$ and $f^-:=\lim_{x\rightarrow\bar{x}^-}f(x)$, it satisfies $f^+\neq f^-$ and  $\max\{f^+,f^-\}\leq r_f$ for some $r_f>0$.}
     at times $t=T_i$ for $i\in\{1,...,N_d\}$,
    $N_d\in\integerNumbers$, $0\leq N_d<\infty$, and
    $0<\inf_{i\in\{1,...,N_d-1\}}(T_{i+1}-T_i)< \infty$ for $N_d>1$.
    In all other moments, the external disturbance function is bounded and has bounded first time derivative, i.e., $\sup_{t\geq0,t\not\in\{T_i\}}|d(t)|<r_d$ and
    $\sup_{t\geq0,t\not\in\{T_i\}}|\dot{d}(t)|<r_{\dot{d}}$
    for some $r_d,r_{\dot{d}}>0$ and
    $i\in\{1,...,N_d\}$.
\end{assumption}
\color{black}

\begin{assumption}
    \label{ass:referenceSignal}
    The reference signal $v_r(t)$ may have a countable number of first-class discontinuity points at times $t=T_i$ for $i\in\{1,...,N_r\}$, $N_r\in\integerNumbers$, $0\leq N_r<\infty$, and
    $0<\inf_{i\in\{1,...,N_r-1\}}(T_{i+1}-T_i)< \infty$ for $N_r>1$.
    There also exists a positive constant $r_{v_r}$, such that $v_{r}(t)$ and its specific time-derivatives satisfy inequality $\sup_{t\geq 0,t\not\in\{T_i\}}\left\{\left|v_r^{(j)}(t)\right|\right\}\leq r_{v_r}$, for $i\in\{1,...,N_r\}$  and $j\in\{0,1,2,3\}$.
    \label{ass:ref2}
\end{assumption}

% \begin{assumption}
%     \label{ass:referenceSignal}
%     There exists a positive constant $r_{v_r}$ such that the reference signal and its specific time-derivatives satisfy inequality $\sup_{t\leq 0}\left\{\left|v_r^{(i)}(t)\right|\right\}\leq r_{v_r}$, for $i\in\{0,1,2,3\}$.
%     \label{ass:ref2}
% \end{assumption}

% \color{black}
% \begin{remark}
%     In the system model~\eqref{eq:DCDCmodel}, $\mu \in [0,1]$ has replaced the discrete input $\mu\in\{0,1\}$, since PWM strategy is used here to generate governing signal from an analog input signal.
% \end{remark}
% \color{black}

% \begin{remark}
%     System model~\eqref{eq:DCDCmodel} is valid only as long as it can be ensured that no saturation occurs in the coil. Otherwise, the inductance $L$ would depend on the current $i_L$ nonlinearly.
% \end{remark}
% \color{black}

% \begin{figure}[t]
% 	\centering
% 		\begin{subfigure}{0.49\textwidth}
% 			\centering
% 			\includegraphics[width=\textwidth]{img/DCDCon.pdf}
% 			\caption{ON case}
% 		\end{subfigure}
% 		\begin{subfigure}{0.49\textwidth}
% 				\centering
% 				\includegraphics[width=\textwidth]{img/DCDCoff.pdf}
% 				\caption{OFF case}
% 		\end{subfigure}
% 		\begin{subfigure}{0.49\textwidth}
% 				\centering
% 				\includegraphics[width=\textwidth]{img/DCDC.pdf}
% 				\caption{\textit{average} case}
% 		\end{subfigure}
% 	\caption{Circuit diagram of the semiconductor realization of the considered DC-DC power converter of ''buck'' type, with diode $VD$ and control switch $VT$.}
% 	\label{fig:DCDC}
% \end{figure}

\addheaders
\subsection{Application of the ADRC principle}
%%%%%%%%%%%%%%%%%%%%%%%%%%%%%%%%%%%%%%%%%
%%%%%%%%%%%%%%%%%%%%%%%%%%%%%%%%%%%%%%%%%
%%%%%%%%%%%%%%%%%%%%%%%%%%%%%%%%%%%%%%%%%

Following the standard ADRC design, system model~\eqref{eq:DCDCmodel} is reformulated, emphasizing its input-output relation
	\begin{equation}
		\frac{d^2v_o(t)}{dt^2}=\underbrace{-\frac{1}{CR}}_{a_1}\frac{dv_o(t)}{dt}\underbrace{-\frac{1}{CL}}_{a_2}v_o(t)+\underbrace{\frac{V_{\text{in}}}{CL}}_{b}\left[\mu(t)+d(t)\right].
		\label{eq:DCDCmodel3}
	\end{equation}
%Now, assuming imperfection of the parameter modeling phase, a following practical model is used:
	%\begin{equation}
		%\frac{d^2v_o(t)}{d^2t}=\underbrace{-\frac{1}{C_0 R_0}}_{d_1}\frac{dv_o(t)}{dt}\underbrace{-\frac{1}{C_0L_0}}_{d_2}v_o(t)+\underbrace{\frac{V_{\text{in}0}}{C_0 L_0}}_{b}\mu(t),
	%\end{equation}

% \textit{Remark 2.} The system model~\eqref{eq:DCDCmodel}, although linear, does not include a range of uncertainties (parametric, structural) and external interferences that are found in an actual device. Also, the linear system description is valid only as long as it can be ensured that no saturation effects occur in the coil. Otherwise, the inductance $L$ would depend on the current $i_L$ nonlinearly.
% Nevertheless, model~\eqref{eq:DCDCmodel} will be used throughout the work for observer and controller synthesis due to its simplicity and the fact that a robust control scheme is considered.

Combining the uncertain (or unknown) terms in~\eqref{eq:DCDCmodel3}, including the imperfect identification of the input gain, results in a following form of the output voltage dynamics
	\begin{equation}
		\ddot{v}_o = \underbrace{a_2 v_o + a_1 \dot{v}_o + b\mu - \hat{b}\mu + bd}_{F(t,\dot{v}_o,v_o,\mu,d)} + \hat{b}\mu = F(\cdot) + \hat{b}\mu,
		\label{eq:DCDCmodel2}
	\end{equation}
where $\hat{b} \neq 0$ is a precise-enough estimate of the input gain $b$ from~\eqref{eq:DCDCmodel3} and $F(\cdot)$ represents the
% matched
\textit{total disturbance} of~\eqref{eq:DCDCmodel2}.

% \color{black}
% \begin{remark}
% \label{rem:2}
%     The practical justification and theoretical validity of aggregating various system terms as part of the total disturbance $F(\cdot)$ in~\eqref{eq:DCDCmodel2} (including control signal an state-dependent variables) has been thoroughly discussed in~\cite{SenChen}.
% \end{remark}
% \color{black}

Since $\referenceVoltage(t)$ and its derivatives may not be known \textit{a priori}, which may lead to possible inability of constructing feedforward signal in~$\controlSignal$, let us reformulate~\eqref{eq:DCDCmodel2} in error-domain
% define a control error $\controlError \triangleq \referenceVoltage - \outputVoltage$ and
%
\begin{align}
    \controlErrorSecondDerivative = \referenceVoltageSecondDerivative - \outputVoltageSecondDerivative = \underbrace{\referenceVoltageSecondDerivative - \totalDisturbance(\cdot)}_{\errorDomainTotalDisturbance(\cdot,\referenceVoltageSecondDerivative)}-\cgpEstimate\controlSignal,
    \label{eq:errorDomainDynamics}
\end{align}
where $e(t)\triangleq v_r(t)-v_o(t)$ is the control error signal and $\errorDomainTotalDisturbance(\cdot)$ is the total disturbance in the error-domain~\cite{errorCEP}. In this article, we utilize a standard form of the ADRC controller
\begin{align}
    \controlSignal = \cgpEstimate^{-1}(\errorDomainTotalDisturbanceEstimate + \stabilizingController),
    \label{eq:generalizedController}
\end{align}
which is constructed to simultaneously compensate the influence of disturbance using the estimated value of total disturbance ($\errorDomainTotalDisturbanceEstimate$) and to stabilize system \eqref{eq:errorDomainDynamics} in a close vicinity of the equilibrium point $\controlError=0$ using the output-feedback stabilizing controller $\stabilizingController$.

\begin{assumption}
\label{ass:boundedControlSignal}
    Stabilizing controller $\mu_0$ has a structure that guarantees the boundedness of $\mu_0(\cdot)$ and $\dot{\mu}_0(\cdot)$. Although this assumption may seem conservative, it is relaxed with the previously introduced Assumptions \ref{ass:CurrentVoltage}, \ref{ass:ExtDistBounded}, and \ref{ass:referenceSignal}.
\end{assumption}
% From~\eqref{eq:generalizedController}, one can notice the key step in ADRC which is fast and accurate estimation of $\errorDomainTotalDisturbanceEstimate$.
\begin{remark}
    Since the disturbance $\errorDomainTotalDisturbance$ and the control variable $\controlSignal$ have equal relative rank,
    % ($r=2$)
    with respect to the voltage $\outputVoltage$ representing the output of the original system (see~\eqref{eq:DCDCmodel}), the total disturbances affecting the error-domain system, described with \eqref{eq:errorDomainDynamics}, meet the so-called matching condition. The specific differences and control solutions for matched and mismatched disturbances have been thoroughly discussed in~\cite{obs-survey2}.
\end{remark}  \color{black}

We will first put the focus on precise and on-line estimation of perturbing term~$F^*(\cdot)$, crucial for proper \textit{active} disturbance rejection. To calculate $\errorDomainTotalDisturbanceEstimate$, we first need to define the extended state $\extendedState = [z_1 \ z_2 \ z_3]^\top \triangleq [\controlError \ \controlErrorDerivative \ \errorDomainTotalDisturbance]^\top\in\extendedStateDomain$, where $\extendedStateDomain\triangleq\{\pmb{x}\in\realNumbers^3 : \module{x}<r_z\}$ for some $r_z\in\realNumbers_{+}$. The dynamics of the state vector $\extendedState$ can be expressed, upon \eqref{eq:errorDomainDynamics}, as a state-space model
\begin{align}
    \begin{cases}
        \extendedStateDerivative = \stateMatrix\extendedState - \inputVector\cgpEstimate\controlSignal + \disturbanceInputVector\errorDomainTotalDisturbanceDerivative, \\
        {\systemOutput = \controlError-\measurementNoise= \outputVector^\top\extendedState - \measurementNoise},
    \end{cases}
    \label{eq:SystemSS}
\end{align}
where $\stateMatrix \triangleq \big[\begin{smallmatrix} \zeroMatrix^{2\times1} & \identityMatrix^{2\times2} \\ 0 & \zeroMatrix^{1\times2} \end{smallmatrix}\big]$, $\inputVector \triangleq [0 \ 1 \ 0]^\top$,  $\outputVector \triangleq [1 \ 0 \ 0]^\top$, and $\disturbanceInputVector\triangleq[0 \ 0 \ 1]^\top$. Given~\eqref{eq:SystemSS}, the output of this system $\systemOutput$ corresponds to the control error $\controlError$ which, according to Assumption~\ref{ass:Output}, is influenced by the measurement noise~$\measurementNoise$.

\begin{remark}
    Control error $\controlError$, together with its derivative $\controlErrorDerivative$ are bounded according to the Assumptions \ref{ass:CurrentVoltage}, \ref{ass:ExtDistBounded}, and \ref{ass:referenceSignal}, and the specific form of the system dynamics \eqref{eq:DCDCmodel}.
\end{remark}
\begin{remark}
    \label{ass:boundedF}
    Under the Assumptions \ref{ass:CurrentVoltage}, \ref{ass:ExtDistBounded} and \ref{ass:referenceSignal},  function $\errorDomainTotalDisturbance(t)$ is continuously differentiable, and thus, there exist bounded continuous functions $\Psi_{\errorDomainTotalDisturbance},\Psi_{\dot{F}^*}$ such that $\sup_{t\geq0}|F^*(t)|< \Psi_{\errorDomainTotalDisturbance}(\controlError,\controlErrorDerivative,\referenceVoltage,\referenceVoltageDerivative,\referenceVoltageSecondDerivative,\mu)$, $\sup_{t\geq0}|\dot{F}^*(t)|< \Psi_{\dot{F}^*}(\controlError,\controlErrorDerivative,\referenceVoltage,\referenceVoltageDerivative,\referenceVoltageSecondDerivative,\referenceVoltageThirdDerivative,\mu,\dot{\mu})$,
    % \begin{equation}
    %     \sup_{t\leq 0}\left\{\left|\errorDomainTotalDisturbance\right|,\left|\frac{\partial\errorDomainTotalDisturbance}{\partial \controlError}\right|,\left|\frac{\partial\errorDomainTotalDisturbance}{\partial \controlErrorDerivative}\right|,\left|\frac{\partial\errorDomainTotalDisturbance}{\partial t}\right|\right\} \leq \Psi_{\errorDomainTotalDisturbance}(\controlError,\controlErrorDerivative,\mu) \nonumber
    % \end{equation}
    for all ${[\controlError \ \controlErrorDerivative]^\top\in \realNumbers^2}$.
    Both practical and theoretical justifications of lumping selected components as parts of $F^*(\cdot)$, including control signal and state-dependent variables, has been thoroughly discussed in~\cite{SenChen}. \color{black}
\end{remark}

% has been thoroughly discussed in~\cite{SenChen}

% \color{black}
% \begin{remark}
%     Assumption~\ref{ass:boundedF} implies that the uncertainty $\errorDomainTotalDisturbance$ and its partial derivatives are bounded if the system states are bounded.
% \end{remark}
% \color{black}

%%%%%%%%%%%%%%%%%%%%%%%%%%%%%%%%%%%%%%%%%
%%%%%%%%%%%%%%%%%%%%%%%%%%%%%%%%%%%%%%%%%
%%%%%%%%%%%%%%%%%%%%%%%%%%%%%%%%%%%%%%%%%
\section{Main result: proposed cascade ESO ADRC}
\label{sect:MainSectCESO}
%%%%%%%%%%%%%%%%%%%%%%%%%%%%%%%%%%%%%%%%%
%%%%%%%%%%%%%%%%%%%%%%%%%%%%%%%%%%%%%%%%%
%%%%%%%%%%%%%%%%%%%%%%%%%%%%%%%%%%%%%%%%%

To calculate the estimated value of extended state vector $\extendedState$, let us now introduce a novel $\cascadeLevel$-level structure of a cascade observer ($\cascadeLevel\in\integerNumbers$ and $\cascadeLevel\geq2$) in a following form
\begin{align}
    \observerStageStateDerivative{1}(t)&=\stateMatrix\observerStageState{1}(t) - \inputVector\cgpEstimate\controlSignal(t)+\observerGainVector{1}\left[\systemOutput(t)-\outputVector^\top\observerStageState{1}(t)\right] \nonumber \\
    \observerStageStateDerivative{i}(t) &= \stateMatrix\observerStageState{i}(t) + \inputVector\left(-\cgpEstimate\controlSignal(t)+\disturbanceInputVector^\top\sum_{j=1}^{i-1}\observerStageState{j}(t)\right) \nonumber \\
    &+\observerGainVector{i}\outputVector^\top\left[\observerStageState{i-1}(t)-\observerStageState{i}(t)\right], \quad i\in\{2,...,\cascadeLevel\},
    \label{eq:iStageObserver}
\end{align}
where $\observerStageState{j}\triangleq[\observerStageStateElement{j,1} \ \observerStageStateElement{j,2} \ \observerStageStateElement{j,3}]^\top\in\realNumbers^3$ is the state of a  particular observer cascade level, $\observerGainVector{j}\triangleq[3\observerStageBandwidth{j} \ 3\observerStageBandwidth{j}^2 \ \observerStageBandwidth{j}^3]^\top\in\realNumbers^3$ is the observer gain vector with design parameter $\observerStageBandwidth{j}\triangleq\observerBandwidthMultiplier^{j-1}\observerStageBandwidth{1}\in\color{black}\realNumbers_{+}\color{black}$ for $\observerBandwidthMultiplier>1, \ \observerStageBandwidth{1}\in\color{black}\realNumbers_{+}\color{black}$, and $j\in\{1,...,\cascadeLevel\}$. The estimate of $\extendedState$, resulting from the observer \eqref{eq:iStageObserver} can be expressed as
\begin{align}
    \extendedStateEstimate = [\extendedStateEstimateElement{1} \ \extendedStateEstimateElement{2} \ \extendedStateEstimateElement{3}]^\top \triangleq \observerStageState{\cascadeLevel} + \disturbanceInputVector\disturbanceInputVector^\top\sum_{j=1}^{\cascadeLevel-1}\observerStageState{j}\in\realNumbers^3.
    \label{eq:extendedStateEstimate}
\end{align}

\begin{remark}
    It is worth noting, that if we reduce the observer to a single level ($\cascadeLevel=1$), we would obtain a standard form of a linear high-gain ESO, as seen in~\cite{GaoLESO}.
    \textcolor{black}{An introduction of the subsequent cascade levels allows us to keep the same observation quality with smaller values of $\observerStageBandwidth{1}$, resulting in a decrease of the measurement noise amplification visible in the state estimates, see~\eqref{eq:iStageObserver}. This effect will be depicted in the upcoming experiments.}
\end{remark}

\begin{figure}[t]
	\centering
		\includegraphics[width=0.49\textwidth]{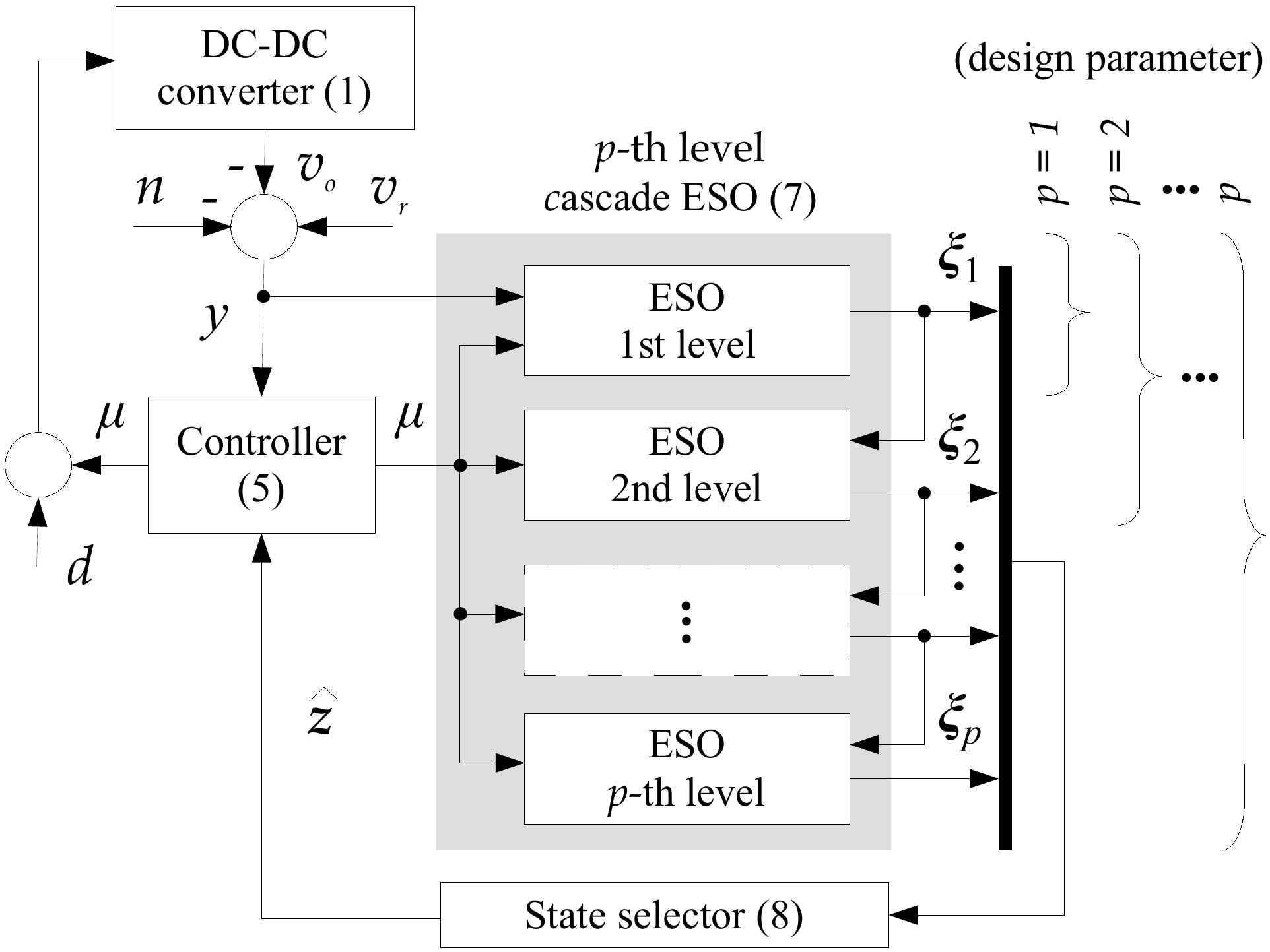}
	\caption{Proposed ADRC with sensor noise suppression via cascade ESO structure for the DC-DC buck power converter.}
	\label{fig:DCDCster}
\end{figure}

\textcolor{black}{The idea of cascade observer structure, proposed in~\eqref{eq:iStageObserver} and illustrated in Fig.~\ref{fig:DCDCster}, is based on a specific choice of the first level observer bandwidth $\observerStageBandwidth{1}$, which should be large enough to guarantee precise estimation of the first element of extended state vector $\extendedState$, and low enough to make the first level of the cascade to act as a low-pass filter for the noise. Latter elements of the extended state vector, i.e. $z_{2}$ and $z_{3}$, usually have faster transients, and thus, are not estimated precisely with the first level observer with a low $\observerStageBandwidth{1}$ value. The consecutive observer levels are introduced to improve the estimation quality of $z_{2}$ and $z_{3}$ using higher observer bandwidths $\omega_{oi}$ ($i>1$) and improve the observation performance by incrementally extending the range of precisely estimated signal frequencies.
\textcolor{black}{The introduction of additional cascade levels of the observer can be interpreted as an attempt to estimate the total disturbance residue, that could not be precisely estimated with the previous cascade levels due to limited bandwidth, and its inclusion in the overall estimate of the extended state vector~\eqref{eq:extendedStateEstimate}.}
The following observer levels are using the state vectors of previous observer levels instead of the measured signal, and thus, result in lower noise amplification than the single-level ESO with high bandwidth. Important part in the utilized cascade observer structure is the state selector~\eqref{eq:extendedStateEstimate}, which defines which estimated state variables (and from which observer level) participate in the controller synthesis~\eqref{eq:generalizedController} to provide improved sensor noise effect suppression.}

% \textcolor{red}{The use of consecutive cascade levels with higher values of particular level bandwidths $\observerStageBandwidth{i}$ allows an incremental augmentation of the precisely estimated frequency range of total disturbance. In other words, a first level cascade estimates total disturbance up to the certain frequency, leaving higher frequency modes filtered. Second level shifts the frequency range a bit, allowing to precisely estimate some of the higher frequencies of total disturbance up to the certain level, etc.}

%   The idea of the cascade observer topology is to estimate the total disturbance observation residue resulting from the previous level with the next cascade level of ESO.
 Having $\extendedStateEstimate$, the application of control action \eqref{eq:generalizedController} to the system \eqref{eq:errorDomainDynamics} results in a following second-order error dynamics
 \begin{align}
     \controlErrorSecondDerivative = \errorDomainTotalDisturbanceObservationError - \stabilizingController,
     \label{eq:errorDomainDynamics2}
 \end{align}
 where $\errorDomainTotalDisturbanceObservationError\triangleq\errorDomainTotalDisturbance - \errorDomainTotalDisturbanceEstimate$ is the final residue of the total disturbance resulting from the imperfect observation of $\errorDomainTotalDisturbance$ by observer~\eqref{eq:iStageObserver}.

%  A more precise derivation of the observer equations can be found in~\cite{lakomy2020cascade}.

A block diagram of the proposed ADRC with cascade ESO for the DC-DC buck power converter is shown in Fig.~\ref{fig:DCDCster}.

\addheaders
% Another advantage of the proposed ADRC scheme is that the same form of control action (xxx) is used, which minimizes potential transition from conventional to the proposed design...

\begin{theorem}
    Under Assumptions~\ref{ass:boundedF}-\ref{ass:boundedControlSignal}, and by taking a stabilizing proportional-derivative controller
    \begin{align}
        \stabilizingController \triangleq \stabilizingControllerProportionalGain\systemOutput + \stabilizingControllerDerivativeGain\extendedStateEstimateElement{2}, \quad \stabilizingControllerProportionalGain,\stabilizingControllerDerivativeGain>0,
        \label{eq:feedbackController}
    \end{align}
    the observation errors of the extended state obtained with the $\cascadeLevel$-level cascade observer, defined as
    \begin{align}
        \extendedStateObservationErrorCascadeLevel{\cascadeLevel} = [\extendedStateObservationErrorCascadeLevelElement{p1} \ \extendedStateObservationErrorCascadeLevelElement{p2} \ \extendedStateObservationErrorCascadeLevelElement{p3}]^\top \triangleq \extendedState - \extendedStateEstimate,
        = \extendedState - \observerStageState{\cascadeLevel} - \disturbanceInputVector\disturbanceInputVector^\top\sum_{j=1}^{\cascadeLevel-1}\observerStageState{j}\in\realNumbers^3,
        \label{eq:observationError}
    \end{align}
    together with the control error $\controlError$, described with the dynamics \eqref{eq:errorDomainDynamics2}, are bounded. In other words
    \begin{align}
         {\displaystyle\color{black}\forall_{t>t_0}\color{black}\forall_{ \observerStageBandwidth{1},\stabilizingControllerParameter>0}  \exists_{\observationErrorUltimateBound,\controlErrorUltimateBound>0}\ls \module{\extendedStateObservationErrorCascadeLevel{\cascadeLevel}(t)}<\observationErrorUltimateBound \wedge  \ls\abs{\controlError(t)}<\controlErrorUltimateBound},
         \label{eq:mainStatement}
    \end{align}
    \color{black}
    where $t_0=\max\{T_{N_d},T_{N_r}\}$ results from Assumptions~\ref{ass:ExtDistBounded} and~\ref{ass:referenceSignal}.\color{black}
\end{theorem}
%
% \begin{remark}
%     \color{red} In the nominal case, when $\measurementNoise\equiv0$, observation and control errors can be made arbitrarily small.
% \end{remark}
%
\begin{remark}
    \label{rem:controllerParameter}
    To keep the notational conciseness of the following theoretical analysis \textcolor{black}{and to reduce the overall number of tuning parameters}, we propose, following~~\cite{GaoLESO}, to tune the stabilizing controller \eqref{eq:feedbackController} with a single parameter $\stabilizingControllerParameter>0$, setting the values of proportional and derivative gains, respectively, as $\stabilizingControllerProportionalGain = \stabilizingControllerParameter^2$ and $ \stabilizingControllerDerivativeGain =   2\stabilizingControllerParameter$. \textcolor{black}{Chosen tuning procedure places the poles of control error dynamics \eqref{eq:errorDomainDynamics2} at value~$-\stabilizingControllerParameter$.}
    % Similar controller parametrization was used in~\cite{GaoLESO}.
\end{remark}

\textit{Proof of Theorem 1.}
% The statement made in the \eqref{eq:mainStatement} will be proven in two steps - for observation subsystem and for control subsystem separately.
The dynamics of the observation error defined for a particular cascade level, i.e. $\extendedStateObservationErrorCascadeLevel{i}\triangleq \extendedState - \observerStageState{i}-\disturbanceInputVector\disturbanceInputVector^\top\sum^{i-1}_{j=1}\observerStageState{j}\in\realNumbers^3 \ \textrm{for} \ i\in\{1,...,\cascadeLevel\}$, can be expressed (after some algebraic transformations) as
\begin{align}
    \extendedStateObservationErrorCascadeLevelDerivative{1} &= (\stateMatrix - \observerGainVector{1}\outputVector^\top)\extendedStateObservationErrorCascadeLevel{1} - \observerGainVector{1}\measurementNoise + \disturbanceInputVector\errorDomainTotalDisturbanceDerivative, \nonumber \\
    \extendedStateObservationErrorCascadeLevelDerivative{i} &= (\stateMatrix-\observerGainVector{i}\outputVector^\top)\extendedStateObservationErrorCascadeLevel{i} + (\observerGainVector{i}\outputVector^\top-\disturbanceInputVector\disturbanceInputVector^\top\observerGainVector{i-1}\outputVector^\top)\extendedStateObservationErrorCascadeLevel{i-1}-\disturbanceInputVector\disturbanceInputVector^\top\observerGainVector{1}\measurementNoise+\disturbanceInputVector\errorDomainTotalDisturbanceDerivative \nonumber \\
    &-\disturbanceInputVector\disturbanceInputVector^\top\sum^{i-2}_{j=1}(\observerGainVector{j}\outputVector^\top-\observerGainVector{j+1}\outputVector^\top)\extendedStateObservationErrorCascadeLevel{j}, \ \textrm{for} \  i\in\{2,...,\cascadeLevel\}.
    \label{eq:observationErrorsDynamics}
\end{align}\addheaders
Equations \eqref{eq:observationErrorsDynamics} allow us to write the dynamics of the aggregated observation error $\extendedStateAggregatedObservationError\triangleq[\extendedStateObservationErrorCascadeLevel{1}^\top \ ... \ \extendedStateObservationErrorCascadeLevel{\cascadeLevel}^\top]^\top\in\realNumbers^{3\cascadeLevel}$ in a form
\begin{align}
    \label{eq:extendedStateAggregatedObservationErrorDerivative}
    \extendedStateAggregatedObservationErrorDerivative = \extendedStateAggregatedStateMatrix\extendedStateAggregatedObservationError + \aggregatedDisturbanceInput\errorDomainTotalDisturbanceDerivative + \aggregatedNoiseVector\measurementNoise,
\end{align}
where matrix $\extendedStateAggregatedStateMatrix$ is lower triangular and its eigenvalues $\eigenvalue{i}\in\{-\observerStageBandwidth{1}, \ -\observerBandwidthMultiplier\observerStageBandwidth{1}, \ ..., \ -\observerBandwidthMultiplier^\cascadeLevel\observerStageBandwidth{1}\}$ for $i\in\{1,...,3\cascadeLevel\}$, vector $\aggregatedDisturbanceInput = [\underbrace{\disturbanceInputVector^\top \ ... \ \disturbanceInputVector^\top}_{\cascadeLevel \ \textrm{times}}]^\top$, and \color{black} $\aggregatedNoiseVector = [\observerGainVector{1}^\top \ \underbrace{\observerGainVector{1}^\top\disturbanceInputVector\disturbanceInputVector^\top \ ... \ \observerGainVector{1}^\top\disturbanceInputVector\disturbanceInputVector^\top}_{\cascadeLevel-1 \ \textrm{times}}]^\top$\color{black}. Introducing the transformation $\extendedStateAggregatedObservationError = \stateTransformationMatrix{\chi}\modifiedExtendedStateAggregatedObservationError$ for $\stateTransformationMatrix{\chi}\triangleq\textrm{blkdiag}\{\observerLevelTransformationMatrix{1},...,\observerLevelTransformationMatrix{p}\}\in\realNumbers^{3\cascadeLevel\times3\cascadeLevel}$ where $\observerLevelTransformationMatrix{i}\triangleq\textrm{diag}\{(\observerBandwidthMultiplier^{i-1}\observerStageBandwidth{1})^{-2}, \ (\observerBandwidthMultiplier^{i-1}\observerStageBandwidth{1})^{-1}, \ 1\}\in\realNumbers^{3\times3}$ for $i\in\{1,...,p\}\in\realNumbers^{\cascadeLevel\times\cascadeLevel}$, we can rewrite \eqref{eq:extendedStateAggregatedObservationErrorDerivative} to a form
\begin{align}
    \modifiedExtendedStateAggregatedObservationErrorDerivative &= \stateTransformationMatrix{\chi}^{-1}\extendedStateAggregatedStateMatrix\stateTransformationMatrix{\chi}\modifiedExtendedStateAggregatedObservationError + \stateTransformationMatrix{\chi}^{-1}\aggregatedDisturbanceInput\errorDomainTotalDisturbanceDerivative + \stateTransformationMatrix{\chi}^{-1}\aggregatedNoiseVector\measurementNoise \nonumber \\
    &= \observerStageBandwidth{1}\transformedObservationErrorStateMatrix \modifiedExtendedStateAggregatedObservationError + \aggregatedDisturbanceInput\errorDomainTotalDisturbanceDerivative + \color{black}\stateTransformationMatrix{\chi}^{-1}\color{black}\aggregatedNoiseVector\measurementNoise,
    \label{eq:transformedObservationErrors}
\end{align}
where $\transformedObservationErrorStateMatrix$ is dependent only on parameter $\alpha$ and its eigenvalues $\eigenvalue{i}\in\{-1, \ -\observerBandwidthMultiplier, ..., \ -\observerBandwidthMultiplier^\cascadeLevel\}$ for $i\in\{1,...,3\cascadeLevel\}$. To conduct a stability analysis of the observation subsystem, let us introduce a Lyapunov function candidate $\observationErrorLyapunovFunction = \modifiedExtendedStateAggregatedObservationError^\top\observationErrorLyapunovEquationSolution\modifiedExtendedStateAggregatedObservationError:\realNumbers^{3\cascadeLevel}\rightarrow\color{black}\realNumbers_{\geq0}\color{black}$ limited by $\minEigenvalue{\observationErrorLyapunovEquationSolution}\module{\modifiedExtendedStateAggregatedObservationError}^2\leq\observationErrorLyapunovFunction\leq\maxEigenvalue{\observationErrorLyapunovEquationSolution}\module{\modifiedExtendedStateAggregatedObservationError}^2$, where $\observationErrorLyapunovEquationSolution\succ0$ is the solution of Lyapunov equation
$ \transformedObservationErrorStateMatrix^\top\observationErrorLyapunovEquationSolution+\observationErrorLyapunovEquationSolution\transformedObservationErrorStateMatrix = -\identityMatrix.$
The derivative of $\observationErrorLyapunovFunction$, based on \eqref{eq:transformedObservationErrors}, can be written down as
\begin{align}
    \observationErrorLyapunovFunctionDerivative &= -\observerStageBandwidth{1}\modifiedExtendedStateAggregatedObservationError^\top\modifiedExtendedStateAggregatedObservationError+2\modifiedExtendedStateAggregatedObservationError^\top\observationErrorLyapunovEquationSolution(\aggregatedDisturbanceInput\errorDomainTotalDisturbanceDerivative+\color{black}\stateTransformationMatrix{\chi}^{-1}\color{black}\aggregatedNoiseVector\measurementNoise) \nonumber \\
    &\leq -\observerStageBandwidth{1}\module{\modifiedExtendedStateAggregatedObservationError}^2 + 2\module{\modifiedExtendedStateAggregatedObservationError}\maxEigenvalue{\observationErrorLyapunovEquationSolution}\sqrt{\cascadeLevel}\left(\abs{\errorDomainTotalDisturbanceDerivative} + \color{black}3\color{black}\observerStageBandwidth{1}^3\abs{\measurementNoise}\right)
\end{align}
and holds
\begin{align}
    \observationErrorLyapunovFunctionDerivative&\leq-(1-\observationErrorMajorizationConstant)\observerStageBandwidth{1}\module{\modifiedExtendedStateAggregatedObservationError} \ \textrm{for} \nonumber \\
    \module{\modifiedExtendedStateAggregatedObservationError}&\geq\frac{2\maxEigenvalue{\observationErrorLyapunovEquationSolution}\sqrt{\cascadeLevel}}{\observerStageBandwidth{1}\observationErrorMajorizationConstant}\abs{\errorDomainTotalDisturbanceDerivative}+\frac{\color{black}6\color{black}\maxEigenvalue{\observationErrorLyapunovEquationSolution}\sqrt{\cascadeLevel}\observerStageBandwidth{1}^2}{\observationErrorMajorizationConstant}\abs{\measurementNoise},
    \label{eq:lowerBoundObservationError}
\end{align}
where $\observationErrorMajorizationConstant\in(0,1)$ is a chosen majorization constant.
The lower bound of $\module{\modifiedExtendedStateAggregatedObservationError}$ is a class $\kappaFunction$ function with respect to the perturturbations $\abs{\errorDomainTotalDisturbanceDerivative}$ and $\abs{\measurementNoise}$, so according to the {Remark~\ref{ass:boundedF} and Assumption~\ref{ass:Output}},  system \eqref{eq:transformedObservationErrors} is input-to-state stable (ISS), and according to \cite{khalil2002}, satisfies
\begin{align}
    \ls\module{\modifiedExtendedStateAggregatedObservationError(t)}&\leq\rho_\chi\frac{2\maxEigenvalue{\observationErrorLyapunovEquationSolution}\sqrt{\cascadeLevel}}{\observerStageBandwidth{1}\observationErrorMajorizationConstant}\Psi_{\dot{F}^*}(\cdot) \nonumber \\
    &+\rho_\chi\frac{\color{black}6\color{black}\maxEigenvalue{\observationErrorLyapunovEquationSolution}\sqrt{\cascadeLevel}\observerStageBandwidth{1}^2}{\observationErrorMajorizationConstant}r_n,
    \label{eq:observationErrorResult}
\end{align}
for $\rho_\chi = \sqrt{\maxEigenvalue{\observationErrorLyapunovEquationSolution}/\minEigenvalue{\observationErrorLyapunovEquationSolution}}$. Since $\maxEigenvalue{\stateTransformationMatrix{\chi}}=\max\{1,(\observationErrorMajorizationConstant^{p-1}\observerStageBandwidth{1})^{-2}\}$ and  $\extendedStateObservationErrorCascadeLevel{p}$ is a subvector of $\extendedStateAggregatedObservationError$, we may write down that $\module{\extendedStateObservationErrorCascadeLevel{p}}\leq\module{\extendedStateAggregatedObservationError}\leq\maxEigenvalue{\stateTransformationMatrix{\chi}}\module{\modifiedExtendedStateAggregatedObservationError}$ and thus that the asymptotic relation
\begin{align}
    \ls\module{\extendedStateObservationErrorCascadeLevel{p}(t)}\leq\maxEigenvalue{\stateTransformationMatrix{\chi}}\ls\module{\modifiedExtendedStateAggregatedObservationError(t)}=:\observationErrorUltimateBound,
\end{align}
which completes the proof of the observer part of~\eqref{eq:mainStatement}.

\begin{remark}
    \label{rem:nominalConditionsObservation}
    Upon the result \eqref{eq:observationErrorResult}, we can see that in the nominal conditions, when $\measurementNoise(t)\equiv0$, the asymptotic relation $\ls\module{\modifiedExtendedStateAggregatedObservationError(t)}\rightarrow0$ as $\observerStageBandwidth{1}\rightarrow\infty$ resulting in the possibility of getting an arbitrarily small value of $\observationErrorUltimateBound$.
\end{remark}

Let us define control error vector $\controlErrorVector=[\controlError \ \controlErrorDerivative]^\top\in\realNumbers^2$. The application of feedback controller \eqref{eq:feedbackController} to dynamics \eqref{eq:errorDomainDynamics2} gives
\begin{align}
    \controlErrorVectorDerivative = \underbrace{\begin{bmatrix} 0 & 1 \\ -\stabilizingControllerParameter^2 & - 2\stabilizingControllerParameter \end{bmatrix}}_{\controlErrorGainMatrix}\controlErrorVector + \underbrace{\begin{bmatrix} 0 & 0 & 0 \\
    0 & 2\stabilizingControllerParameter & 1\end{bmatrix}}_{\controlErrorPerturbationMatrix}\extendedStateObservationErrorCascadeLevel{p} - \underbrace{\begin{bmatrix}0 \\ \stabilizingControllerParameter^2\end{bmatrix}}_{\measurementNoiseInputVector}\measurementNoise,
\end{align}
which can be transformed with substitution $\controlErrorVector = \stateTransformationMatrix{\varepsilon}\transformedControlError$, where  $\stateTransformationMatrix{\varepsilon} \triangleq \textrm{diag}\{\stabilizingControllerParameter^{-1}, \ 1\}$, into
\begin{align}
    \transformedControlErrorDerivative &= \stateTransformationMatrix{\varepsilon}^{-1}\controlErrorGainMatrix\stateTransformationMatrix{\varepsilon}\transformedControlError + \stateTransformationMatrix{\varepsilon}^{-1}\controlErrorPerturbationMatrix\extendedStateObservationErrorCascadeLevel{\cascadeLevel} - \stateTransformationMatrix{\varepsilon}^{-1}\measurementNoiseInputVector\measurementNoise \nonumber \\
    &= \stabilizingControllerParameter\underbrace{\begin{bmatrix} 0 & 1 \\ -1 & -2\end{bmatrix}}_{\transformedControlErrorStateMatrix}\transformedControlError+\controlErrorPerturbationMatrix\extendedStateObservationErrorCascadeLevel{\cascadeLevel}-\measurementNoiseInputVector\measurementNoise.
    \label{eq:transformedControlError}
\end{align}
Let us now introduce a Lyapunov function candidate $\transformedControlErrorLyapunovFunction = \transformedControlError^\top\transformedControlErrorLyapunovEquationSolution\transformedControlError:\realNumbers^2\rightarrow\color{black}\realNumbers_{\geq0}\color{black}$ limited by $\minEigenvalue{\transformedControlErrorLyapunovEquationSolution}\module{\transformedControlError}\leq\transformedControlErrorLyapunovFunction(\transformedControlError)\leq\maxEigenvalue{\transformedControlErrorLyapunovEquationSolution}\module{\transformedControlError}$, where $\transformedControlErrorLyapunovEquationSolution\succ0$ is the solution of Lyapunov equation
$\transformedControlErrorStateMatrix^\top\transformedControlErrorLyapunovEquationSolution+\transformedControlErrorLyapunovEquationSolution\transformedControlErrorStateMatrix=-\identityMatrix.$
The derivative
\begin{align}
    \transformedControlErrorLyapunovFunctionDerivative &= -\stabilizingControllerParameter\transformedControlError^\top\transformedControlError+2\transformedControlError^\top\transformedControlErrorLyapunovEquationSolution\controlErrorPerturbationMatrix\extendedStateObservationErrorCascadeLevel{\cascadeLevel}-2\transformedControlError^\top\transformedControlErrorLyapunovEquationSolution\measurementNoiseInputVector\measurementNoise \nonumber \\
    &\leq -\stabilizingControllerParameter\module{\transformedControlError}^2+\module{\transformedControlError}\maxEigenvalue{\transformedControlErrorLyapunovEquationSolution}\left[\controlErrorPerturbationMatrixMaxValue\module{\extendedStateObservationErrorCascadeLevel{\cascadeLevel}}+\stabilizingControllerParameter^2\abs{\measurementNoise}\right],
\end{align}
where $\controlErrorPerturbationMatrixMaxValue=\max\{1,2\stabilizingControllerParameter\}$,  holds
\begin{align}
    \transformedControlErrorLyapunovFunctionDerivative&\leq-(1-\transformedControlErrorMajorizationConstant)\stabilizingControllerParameter\module{\transformedControlError}^2 \ \textrm{for} \nonumber \\
    \module{\transformedControlError}&\geq\frac{2\maxEigenvalue{\transformedControlErrorLyapunovEquationSolution}}{\transformedControlErrorMajorizationConstant\stabilizingControllerParameter}\left[\controlErrorPerturbationMatrixMaxValue\module{\extendedStateObservationErrorCascadeLevel{\cascadeLevel}}+\stabilizingControllerParameter^2\abs{\measurementNoise}\right]
\end{align}
The lower boundary of $\module{\transformedControlError}$ is class $\kappaFunction$ with respect to arguments $\module{\extendedStateObservationErrorCascadeLevel{\cascadeLevel}}$ and $|\measurementNoise|$. According to the {Remark~\ref{ass:boundedF}, Assumption~\ref{ass:Output}} and result \eqref{eq:lowerBoundObservationError}, system \eqref{eq:transformedControlError} is ISS and satisfies
\begin{align}
    &\ls\module{\transformedControlError(t)}\leq\rho_\varepsilon\frac{2\maxEigenvalue{\transformedControlErrorLyapunovEquationSolution}}{\transformedControlErrorMajorizationConstant\stabilizingControllerParameter}\left[\controlErrorPerturbationMatrixMaxValue\ls\module{\extendedStateObservationErrorCascadeLevel{\cascadeLevel}(t)}+\stabilizingControllerParameter^2r_n\right] \nonumber \\
    &\leq \rho_\varepsilon\frac{2\maxEigenvalue{\transformedControlErrorLyapunovEquationSolution}}{\transformedControlErrorMajorizationConstant\stabilizingControllerParameter}\Bigg[\rho_\chi\frac{2\controlErrorPerturbationMatrixMaxValue\maxEigenvalue{\observationErrorLyapunovEquationSolution}\sqrt{\cascadeLevel}}{\observerStageBandwidth{1}\observationErrorMajorizationConstant}\Psi_{\dot{F}^*}(\cdot) \nonumber \\
    &+\left(\rho_\chi\frac{2\controlErrorPerturbationMatrixMaxValue\maxEigenvalue{\observationErrorLyapunovEquationSolution}\sqrt{\cascadeLevel}\observerStageBandwidth{1}^2}{\observationErrorMajorizationConstant}+\stabilizingControllerParameter^2\right)r_n\Bigg],
\end{align}
where $\rho_\varepsilon = \sqrt{\maxEigenvalue{\transformedControlErrorLyapunovEquationSolution}/\minEigenvalue{\transformedControlErrorLyapunovEquationSolution}}$.
According to transformation between original control error vector $\controlErrorVector$ and the transformed $\transformedControlError$, we write $\module{\controlErrorVectorDerivative}\leq\max\{k^{-1},1\}\module{\transformedControlError}=:m_k\module{\transformedControlError}$ and thus
\begin{align}
    &\ls\module{\controlErrorVector(t)}\leq m_k\rho_\varepsilon\frac{2\maxEigenvalue{\transformedControlErrorLyapunovEquationSolution}}{\transformedControlErrorMajorizationConstant\stabilizingControllerParameter}\Bigg[\rho_\chi\frac{4\maxEigenvalue{\observationErrorLyapunovEquationSolution}\sqrt{\cascadeLevel}}{\observerStageBandwidth{1}\observationErrorMajorizationConstant}\Psi_{\dot{F}^*}(\cdot) \nonumber \\
    &+\left(\rho_\chi\frac{4\maxEigenvalue{\observationErrorLyapunovEquationSolution}\sqrt{\cascadeLevel}\observerStageBandwidth{1}^2}{\observationErrorMajorizationConstant}+\max\{k^{-1},1\}\stabilizingControllerParameter^2\right)r_n\Bigg]=:\controlErrorUltimateBound,
    \label{eq:controlErrorResult}
\end{align}
which completes the proof of Theorem 1.\hspace{1em plus 1fill}\IEEEQEDhere
\begin{remark}
    Similarly to the comment made in Remark~\ref{rem:nominalConditionsObservation}, in the case of $\measurementNoise(t)\equiv0$ and upon the result \eqref{eq:controlErrorResult}, we can say that $\ls\module{\controlErrorVector(t)}\rightarrow0$ as $\observerStageBandwidth{1}\rightarrow\infty \ \vee \ \stabilizingControllerParameter\rightarrow\infty$, making it possible to get an arbitrarily small value of $\controlErrorUltimateBound$.
\end{remark}

\begin{remark}
    Upon the result \eqref{eq:controlErrorResult}, we may observe that the increasing gains of both observer and controller are amplifying measurement noise, thus, it is not recommended to use extremely high values of $\observerStageBandwidth{1}$ and $\stabilizingControllerParameter$ in practice.
\end{remark}

% \begin{align}
%     \observerStageObservationErrorDerivative{i}
%     &= \observationErrorStateMatrix{i}\observerStageObservationError{i} + {\residualObservationErrorMatrix{i,i-1}}\observerStageObservationError{i-1} +\inputVector{n+1}^\top\totalDisturbanceDerivative+\sum_{j=1}^{i-2}\residualObservationErrorMatrix{i,j}\observerStageObservationError{j}
% \end{align}

% Based on the \color{red} simulation/experimental \color{black} results, the relative value of $\observerStageBandwidth{1}$ chosen for conventional ESO is higher and thus visibly amplifies the sensor noise more .

% in the case of proposed ESO these values can be chosen smaller relative to the conventional ESO, which reduces the level of noise amplification, while

% \begin{figure}[t]
% 	\centering
% 	\includegraphics[width=0.49\textwidth]{img/example.pdf}
% 	\caption{SIMULATION: First results for $\omega_{o1} = 4500$ for regular ESO $p=1$, and $\omega_{o1} = 500$ for ESO $p=3$ with $\alpha=3$.}
% 	\label{fig:primaryResults}
% \end{figure}

%%%%%%%%%%%%%%%%%%%%%%%%%%%%%%%%%%%%%%%%%
%%%%%%%%%%%%%%%%%%%%%%%%%%%%%%%%%%%%%%%%%
%%%%%%%%%%%%%%%%%%%%%%%%%%%%%%%%%%%%%%%%%
\section{Hardware experiment}
%%%%%%%%%%%%%%%%%%%%%%%%%%%%%%%%%%%%%%%%%
%%%%%%%%%%%%%%%%%%%%%%%%%%%%%%%%%%%%%%%%%
%%%%%%%%%%%%%%%%%%%%%%%%%%%%%%%%%%%%%%%%%

%%%%%%%%%%%%%%%%%%%%%%%%%%%%%%%%%%%%%%%%%
\subsection{Testbed description}
\label{lab:TesbedDescr}
%%%%%%%%%%%%%%%%%%%%%%%%%%%%%%%%%%%%%%%%%

\begin{figure}[t]
	\centering
	\includegraphics[width=0.4\textwidth]{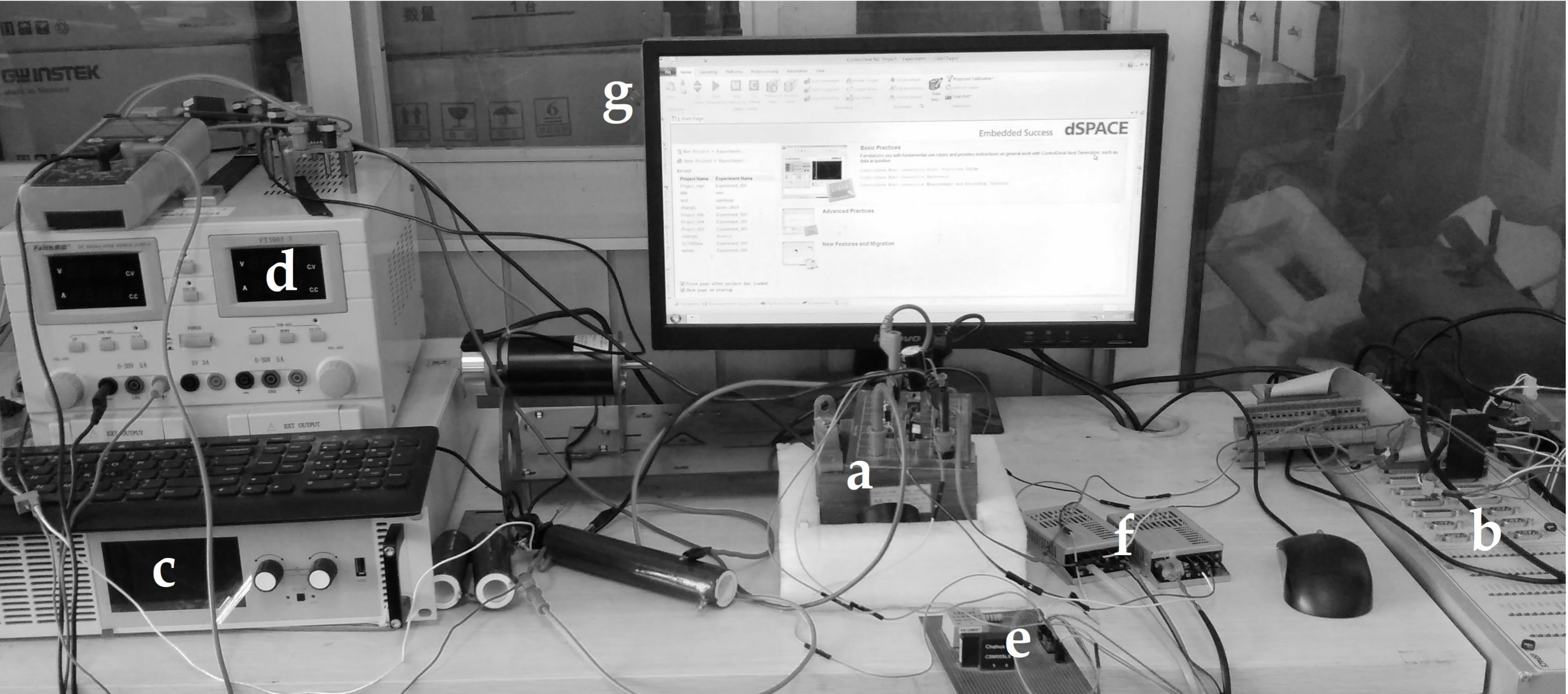}
	\caption{Laboratory setup, with a - buck converter, b - dSPACE controller, c - input voltage, d - oscilloscope, e - voltage sensor, f - A/D converters, and g - PC with control software.}
	\label{fig:platformConfa}
\end{figure}

The experimental setup used for the study is seen in Fig.~\ref{fig:platformConfa}. The output voltage was measured by a Hall effect-based sensor and converted through a 16-bit A/D converter in the dSPACE platform. The output was recorded by a digital oscilloscope and dedicated PC-based software. The sampling period was set to $T_s=10^4$Hz. The physical parameters of the DC-DC converter, described with~\eqref{eq:DCDCmodel}, were $V_{\text{in}}=20$V, $L=0.01$H, $C=0.001$F, and $R=50\Omega$. This allowed to straightforwardly calculate the system gain in~\eqref{eq:DCDCmodel2} as $\hat{b}=V_{\text{in}}/(CL)=2\times 10^6$. The tested control algorithm was first implemented in a Matlab/Simulink-based model, from which a C code program was generated and run on the dSPACE controller in real-time.

\addheaders

% \begin{table}[b]
% \renewcommand{\arraystretch}{1.3}
% \caption{Used bandwidth parameterization of CESOs.}
%  \centering
%  \label{tab:obsbandparamSPECIFIC}
%  \resizebox{\columnwidth}{!}{
% \begin{tabular}{cccc}
% \hline\hline
% \diagbox{Bandwidth}{Cascade level}        & $p=1$             & $p=2$                 & $p=3$                  \\ \hline
% 1st level ESO  & $\omega_{o1}$ & $\frac{\omega_{o1}}{\alpha}$     & $\frac{\omega_{o1}}{\alpha^2}$ \\
% 2nd level ESO & --            & $\omega_{o1}$ & $\frac{\omega_{o1}}{\alpha}$  \\
% 3rd level ESO  & --            & --                & $\omega_{o1}$      \\ \hline\hline
% \end{tabular}}
% \end{table}

\begin{table}[b]
\renewcommand{\arraystretch}{1.3}
\caption{Used bandwidth parameterization of CESOs.}
 \centering
 \label{tab:obsbandparamSPECIFIC}
 \resizebox{\columnwidth}{!}{
\begin{tabular}{cccc}
\hline\hline
\diagbox{Bandwidth}{Cascade level}        & $p=1$             & $p=2$                 & $p=3$                  \\ \hline
1st level ESO ($\omega_{o1}$)  & $\lambda$ & $\frac{\lambda}{\alpha}$     & $\frac{\lambda}{\alpha^2}$ \\
2nd level ESO ($\omega_{o2}$) & --            & $\lambda$ & $\frac{\lambda}{\alpha}$  \\
3rd level ESO ($\omega_{o3}$)  & --            & --                & $\lambda$      \\ \hline\hline
\end{tabular}}
\end{table}

\begin{figure}[t]
	\centering
	\includegraphics[width=0.485\textwidth]{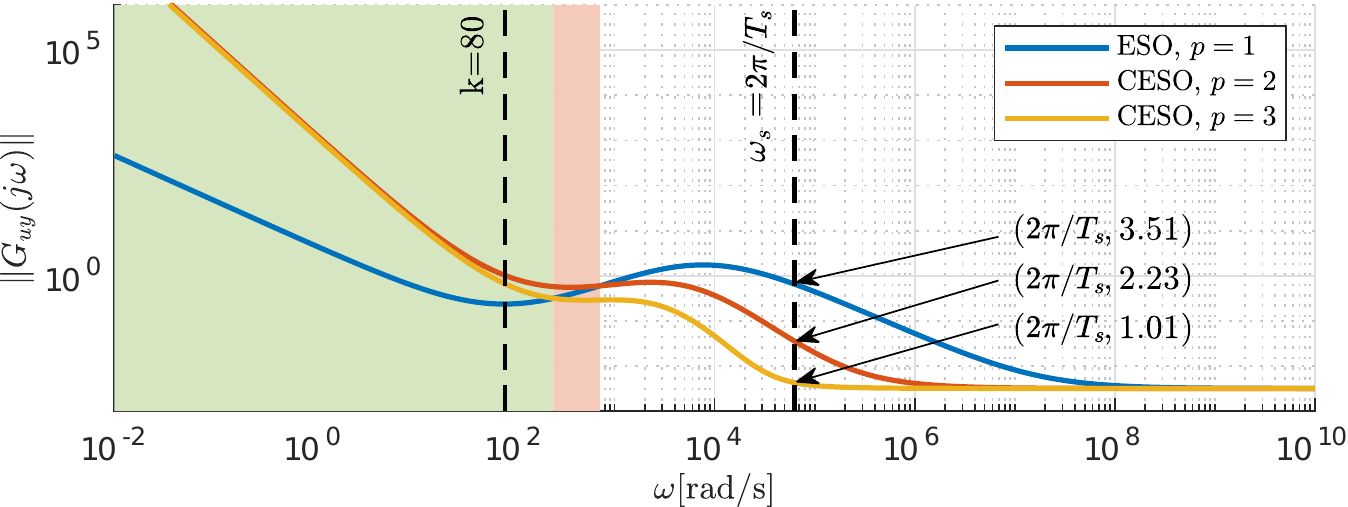}
	\caption{\color{black} Bode diagram representing the module of $G_{uy}(j\omega)$. \color{black}}
	\label{fig:gue}
\end{figure}

\color{black}
Considering the above parameters of the utilized testbed and the controller/observer structures introduced in \eqref{eq:generalizedController},  \eqref{eq:feedbackController}, and \eqref{eq:iStageObserver}, we can derive the transfer-function-based relation
\begin{align}
    U(j\omega)=G_{uy}(j\omega)\underbrace{\left[E(j\omega)-N(j\omega)\right]}_{Y(j\omega)},
\end{align}
where $U(j\omega)$, $E(j\omega)$, $N(j\omega)$, and $Y(j\omega)$ correspond respectively to signals $\controlSignal(t)$, $\controlError(t)$, $n(t)$, and $y(t)$ after Laplace transformation. The amplitude Bode diagram of $G_{uy}(j\omega)$, obtained for the observer levels $p\in\{1,2,3\}$ and tuned with the nominal parameters utilized in the experiment, is presented in Fig.~\ref{fig:gue}. The vertical dashed lines represent the chosen controller bandwidth $k$, which is the range we expect the closed-loop system to operate in, and the experiment sampling frequency $\omega_s$.
The green area represents the frequency range, where CESO ($p=2$ and $p=3$) should react more rapidly than the standard ESO, and red area is the range where only CESO $p=2$ should provide quicker response with respect to control errors. The points at the intersection of $\omega_s$ and observers graphs indicate the amplification factors of high frequency signals (e.g. measurement noise) within signal $\controlSignal(t)$. Consequently, in the following experiments, we can expect the measurement noise to be least amplified in CESO $p=3$, followed by CESO $p=2$, and finally in standard, single ESO.

\color{black}

% The converter is controlled using a PWM gate drive, which compares the duty ratio (control signal) with a triangle wave signal, which results in the generation of the PWM signal.

%%%%%%%%%%%%%%%%%%%%%%%%%%%%%%%%%%%%%%%%%
\subsection{\textcolor{black}{Test methodology}}
\label{sec:Testmethoddd}
%%%%%%%%%%%%%%%%%%%%%%%%%%%%%%%%%%%%%%%%%

The following experiments were conducted to test the ADRC scheme with the proposed cascade ESO (CESO):
    \begin{itemize}
        \item [E1:] Comparison with standard ESO (i.e. CESO with $p=1$).
        \item [E2:] Influence of parameters $\omega_{o1}$ (E2a), $k$ (E2b), and $\alpha$ (E2c).
        \item [E3:] \color{black} Impact of a low-pass filter (LPF) at the converter output.\color{black}
    \end{itemize}

% \begin{remark}
% Please note that a standard, single ESO is synonymous with the CESO with cascade level $p=1$.
% \end{remark}

\begin{table}[b]
\renewcommand{\arraystretch}{1.425}
\caption{Integral quality criteria for experiment E1.}
 \centering
 \label{tab:qualityCriteria2}
%  \vspace{-0.2cm}
\resizebox{\columnwidth}{!}{
\begin{tabular}{c c c c}
\hline\hline \\[-4mm]
% \cline{1-3}
\multirow{2}{*}{Observer type} & \multicolumn{3}{c}{Criterion} \\ \cline{2-4}
 & $\int|e(t)|dt$ & $\int|\controlSignal(t)|dt$ & $\int|\dot{\controlSignal}(t)|dt$ \\ \hline
Standard ESO ($p=1$) & 0.2310 & 0.5368 & 315.58 \\
Cascade ESO ($p=2$) & 0.0467 & 0.5496 & 113.23 \\
Cascade ESO ($p=3$) & 0.0381 & 0.5545 & 29.11 \\ \hline\hline
\end{tabular}
}
\end{table}

%\color{cyan} Robustness (against uncertainty in $R_0$ and $V_{\text{in}}$). ExpXXX: Porownanie ze zwyklym low-pass filter pierwszego rzedu o roznym poziomie odciecia Inne: ? rozne sygn. ref.? \color{black}

\begin{figure}[t]
	\centering
	\vspace{0.30cm} % Raf: dodalem, aby wyrownac z Fig.6
	\includegraphics[width=0.49\textwidth]{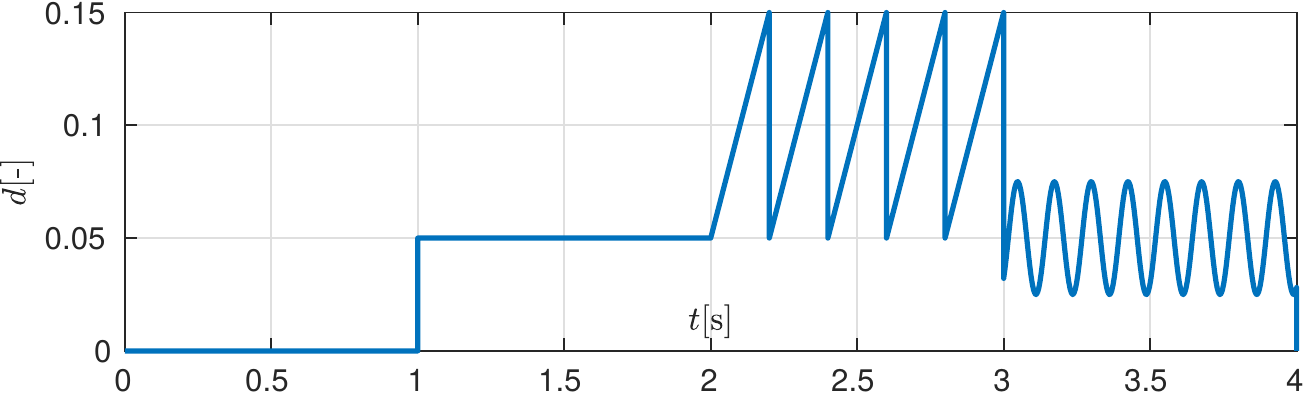}
	\caption{External disturbance applied in all experiments.}
	\label{fig:extdist}
\end{figure}

\textcolor{black}{The control objective was to track a smooth voltage trajectory $v_r(t)$ despite the presence of a varying input-additive external disturbance shown in Fig.~\ref{fig:extdist}. Such disturbance signal is used here to test the robustness of the considered controllers against different types of disturbances within one experimental run. This specific shape of user-injected external disturbance signal would not appear outside of a laboratory environment, however, the character of disturbances designed in specific time intervals can be found in certain applications (e.g.~\cite{TransCircuits2}). The reference trajectory was designed as a filtered and biased square signal with bias equal to 7V, amplitude of square signal equal to 6V, and period 1s. The filtering transfer function applied to the square signal was $G_f(s) = \frac{4}{0.025s^2+0.6s+4}$.
Although the most common control task in the control of buck converters is a set-point stabilization, trajectory following of the output voltage can be occasionally seen in the literature, e.g.,~\cite{SinRefVoltageDCconve}.
%Although this type of reference signal is occasionally used in buck converter control, it is not commonly seen in industrial practice.
Here, we consider a filtered piece-wise constant reference to reach the desired level of the output voltage and avoid observer peaking caused by the discontinuities in~$v_r(t)$.}

%%%%%%%%%%%%%%%%%%%%%%%%%%%%%%%%%%%%%%%%%
\subsection{\textcolor{black}{Experimental results}}
\label{sec:Results}
%%%%%%%%%%%%%%%%%%%%%%%%%%%%%%%%%%%%%%%%%

\begin{figure}[t]
	\centering
	\includegraphics[width=0.49\textwidth]{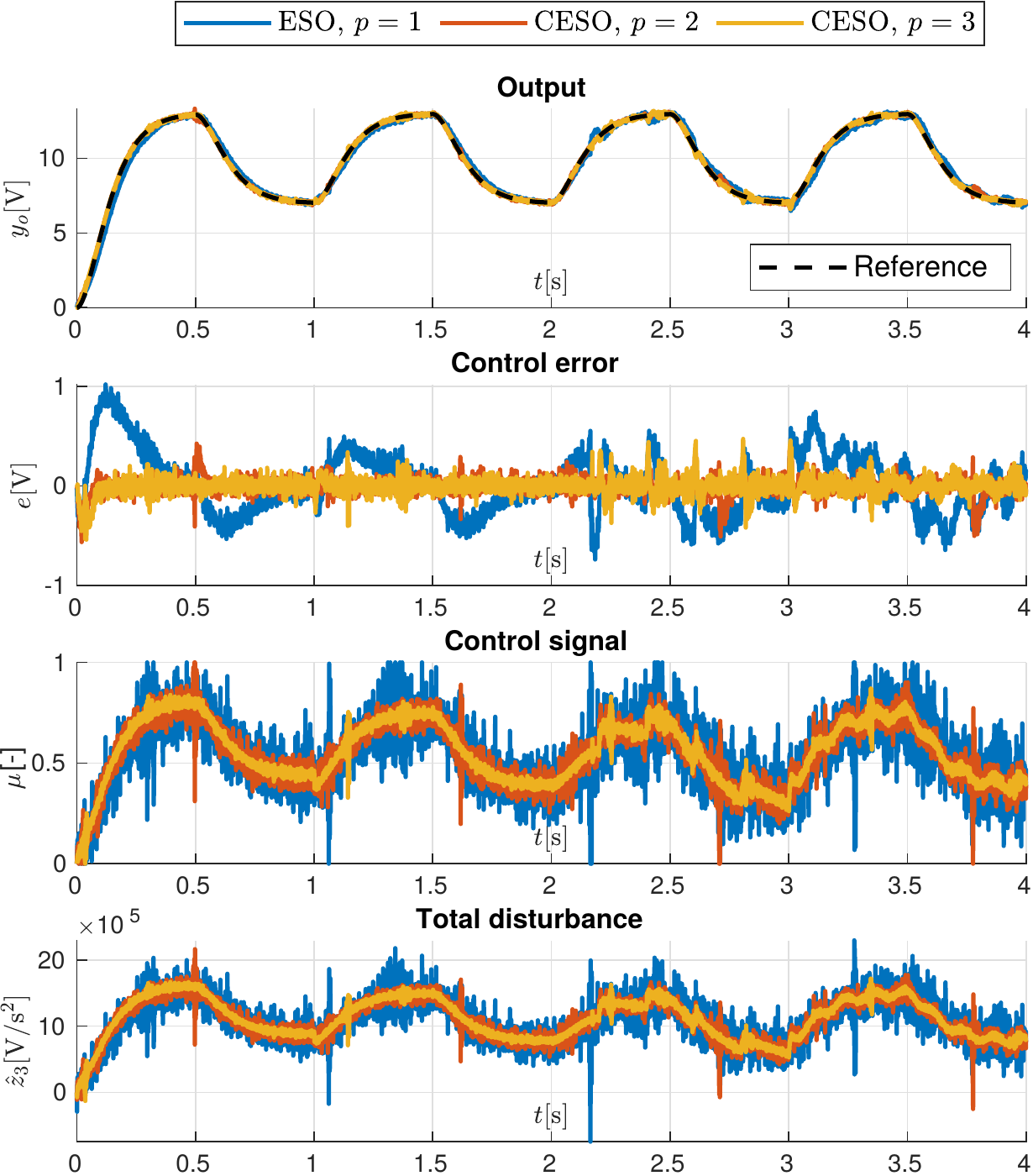}
% 	\caption{Results of E1 - comparison between CESOs with different $p$ value, $\omega_{o, max} = 3600$.}
\caption{Results of experiment E1.}
	\label{fig:E1}
\end{figure}

The results of E1 are gathered in Fig.~\ref{fig:E1}. The observer bandwidth for the standard ESO ($p=1$) was $\observerStageBandwidth{1}=3600$rad/s, which was close to the maximum that could be obtained for a $10$kHz sampling without observing any undesirable effects. For the comparison, only CESOs with $p=2$ and $p=3$ levels were utilized to maintain legibility of the results while not loosing their generality. In order to provide a systematic tuning methodology across tested observers, bandwidths of the CESOs were parameterized and set according to Table~\ref{tab:obsbandparamSPECIFIC} with $\alpha=3$ \color{black} and $\lambda=3600$rad/s\color{black}. The controller gains from~\eqref{eq:feedbackController} were set to $\stabilizingControllerProportionalGain=6400$ and $\stabilizingControllerDerivativeGain=160$ in each case, which corresponds to the controller bandwidth $\stabilizingControllerParameter=80$, introduced in Remark~\ref{rem:controllerParameter}.

One can notice from Fig.~\ref{fig:E1} that, with the applied tuning methodology, all the tested controllers have realized the given task, however the standard ESO ($p=1$) provided the worst performance in terms of tracking accuracy and noise suppression. On the other hand, with the increase of cascade level $p$ in CESO, better performance was achieved. This observation is supported with the calculated integral quality indices in Table~\ref{tab:qualityCriteria2}. Besides the improvement of control error performance, the transfer of sensor noise into the control signal has decreased with the increase of parameter $p$ thanks to the lower values of $\omega_{o1}$ related to the first level of CESO. This result is supported with the values of $\int|\dot{u}(t)|dt$ criterion in Table~\ref{tab:qualityCriteria2}, which represents the impact of rapid fluctuations of the control signal, mostly caused by the amplified noise.

\color{black}
The initial premises formulated upon Fig.~\ref{fig:gue} have been confirmed with the results in Fig.~\ref{fig:E1}. As expected, the control signal with the lowest content of noise was obtained for CESO $p=3$, then CESO $p=2$, and finally the standard ESO.
% The visual relation of the noise level visible in the plot considering $e[V]$, and the noise content in $\mu[-]$ is

% The  associated with the proposed cascade observer structure and obtained in the experiment correspond to the ones calculated in the frequency domain (cf. Fig.~\ref{fig:gue}).
\color{black}

\begin{figure}[p]
    \centering
  \subfloat[standard ESO ($p=1$)]{%
       \includegraphics[width=0.456\textwidth]{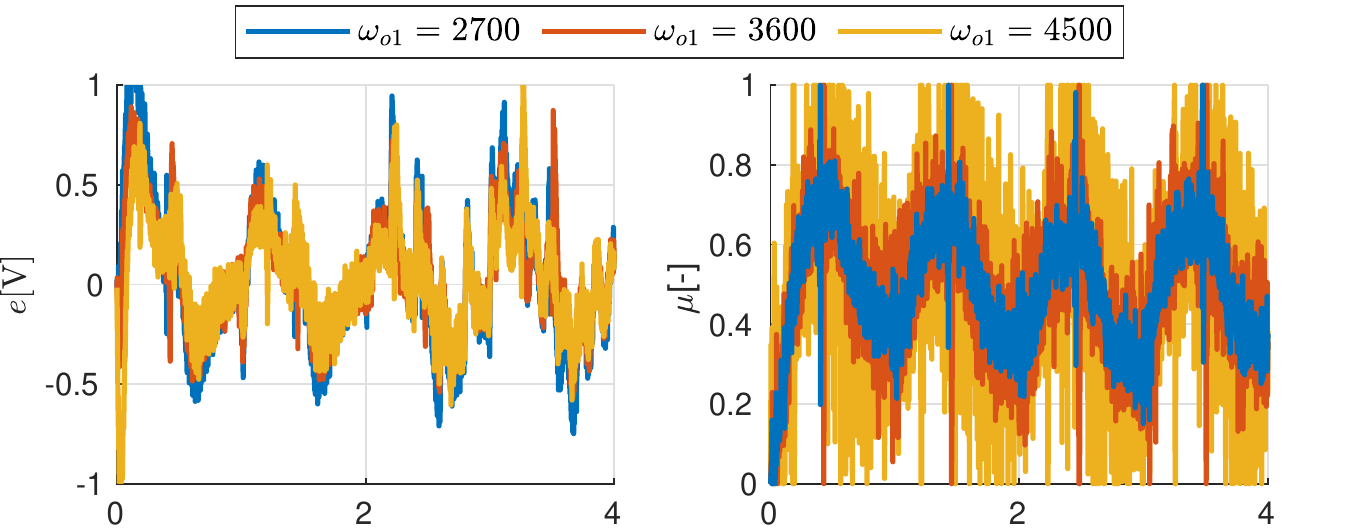}}
    \\
  \subfloat[CESO ($p=2$)]{%
        \includegraphics[width=0.456\textwidth]{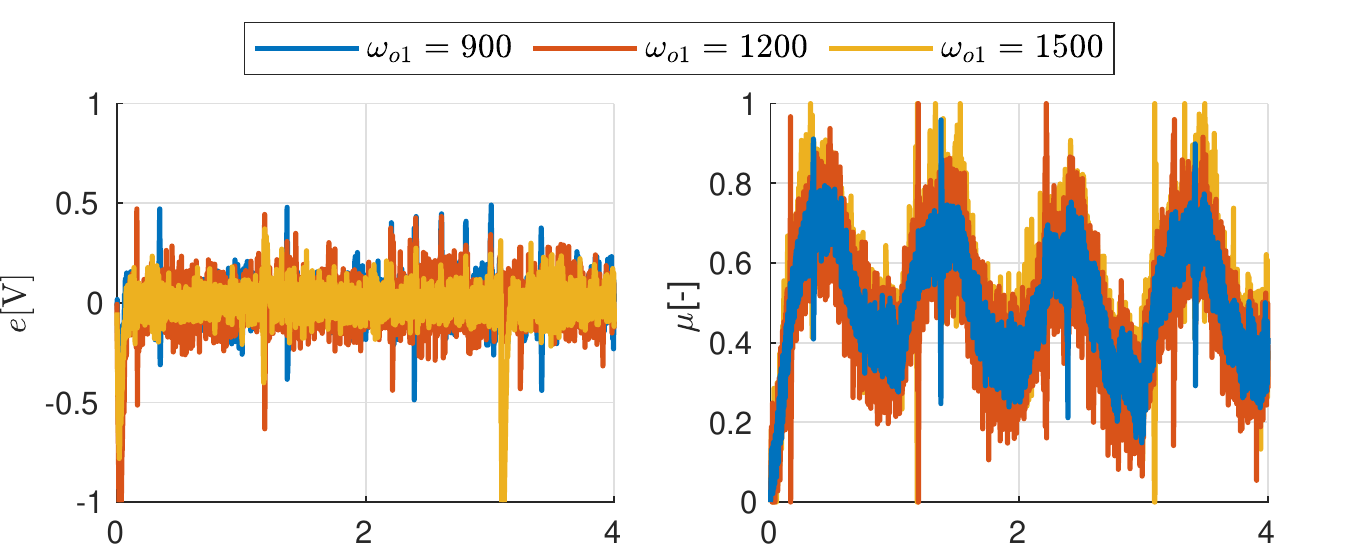}}
    \\
  \subfloat[CESO ($p=3$)]{%
        \includegraphics[width=0.456\textwidth]{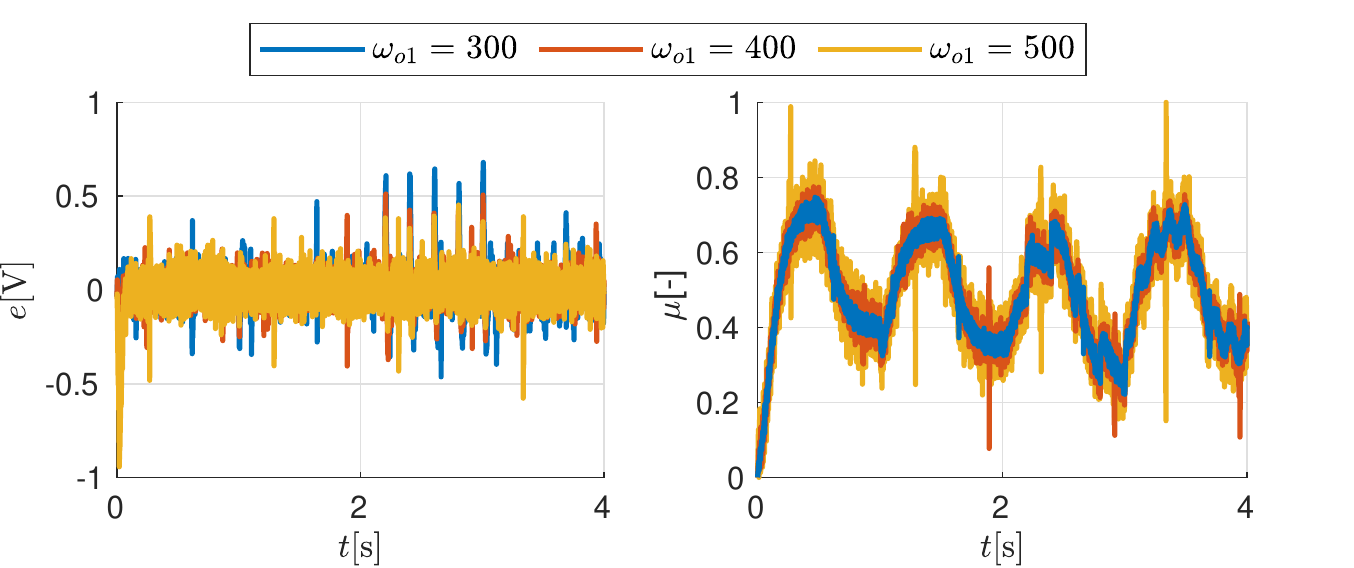}}
  \caption{Results of experiment E2a.}
  \label{figExp2a}
\end{figure}

\begin{figure}[p]
    \centering
  \subfloat[standard ESO ($p=1$)]{%
       \includegraphics[width=0.456\textwidth]{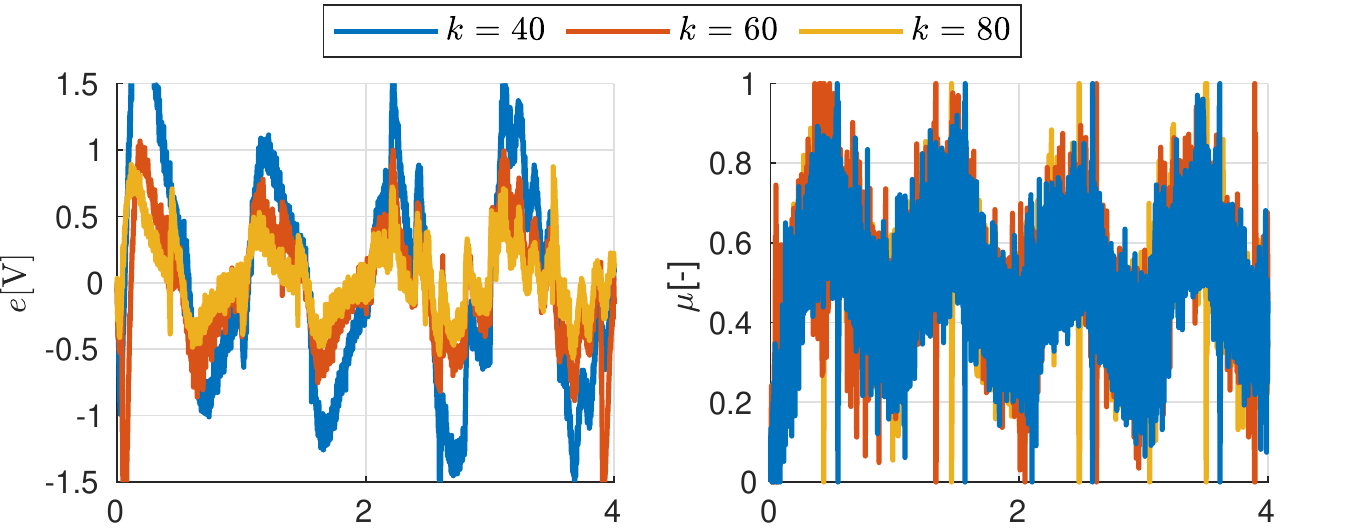}}
    \\
  \subfloat[CESO ($p=2$)]{%
        \includegraphics[width=0.456\textwidth]{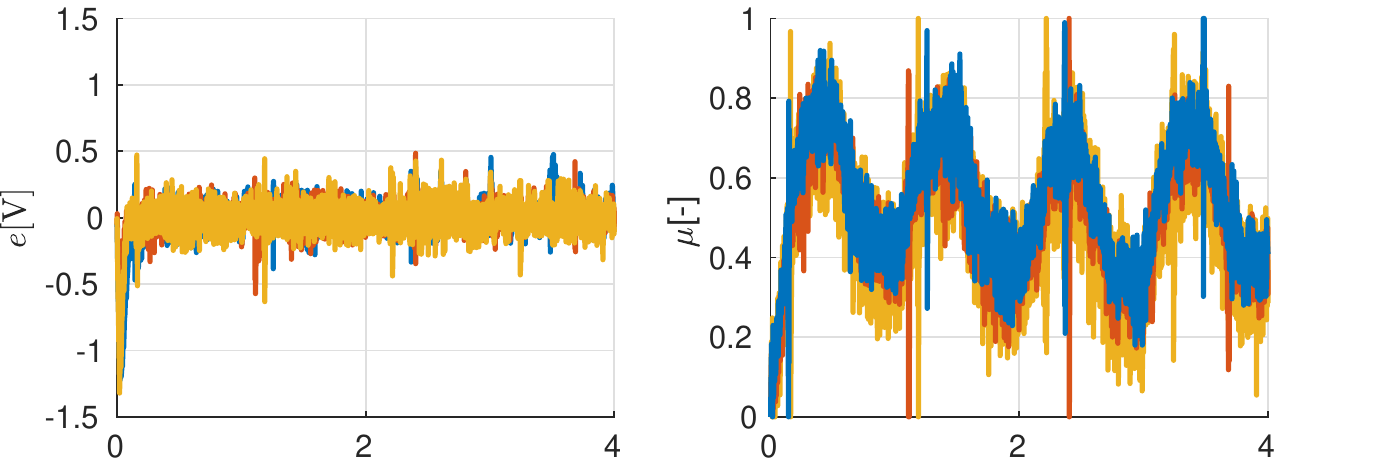}}
    \\
  \subfloat[CESO ($p=3$)]{%
        \includegraphics[width=0.456\textwidth]{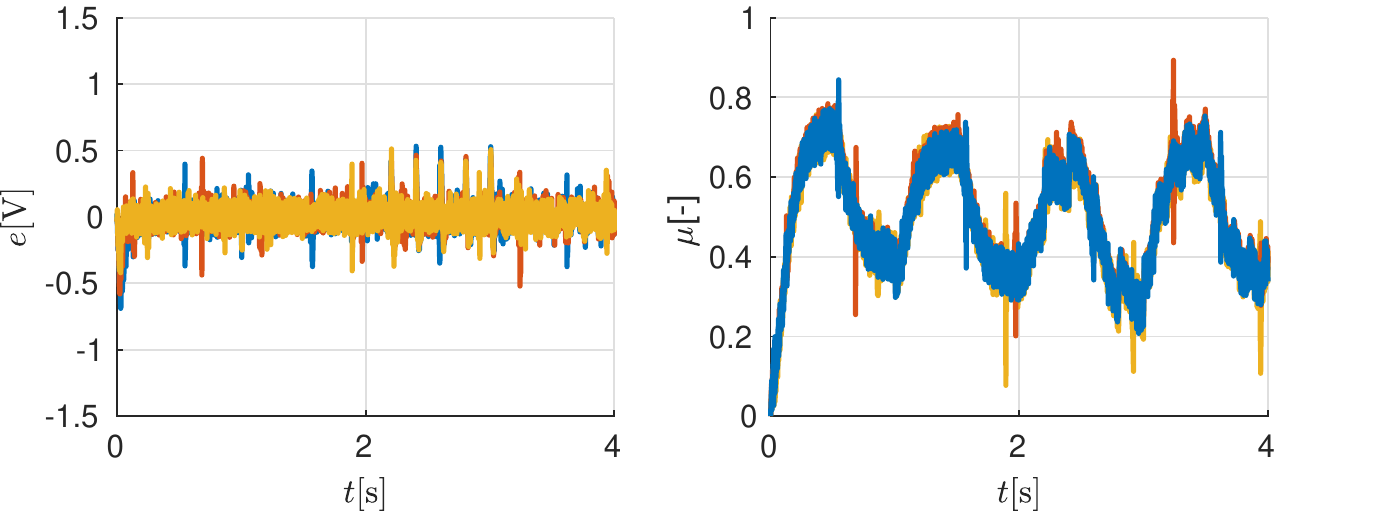}}
  \caption{Results of experiment E2b.}
  \label{figExp2b}
\end{figure}

\begin{figure}[t]
    \centering
  \subfloat[CESO $p=2$ \label{figExp2c1}]{%
       \includegraphics[width=0.456\textwidth]{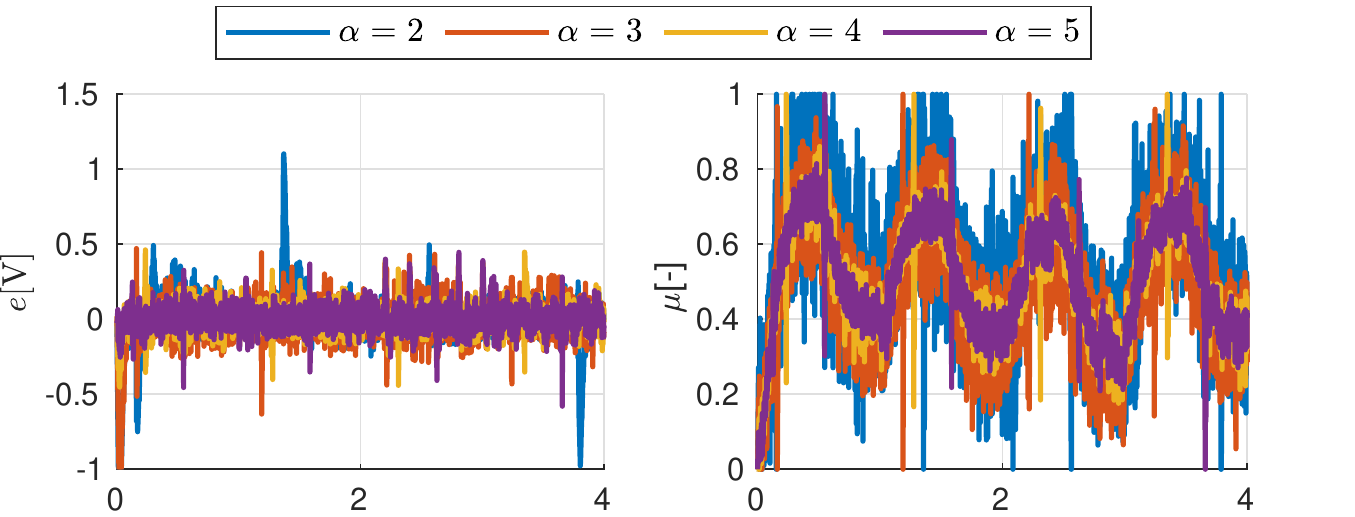}}
    \\
  \subfloat[CESO $p=3$\label{figExp2c2}]{%
        \includegraphics[width=0.456\textwidth]{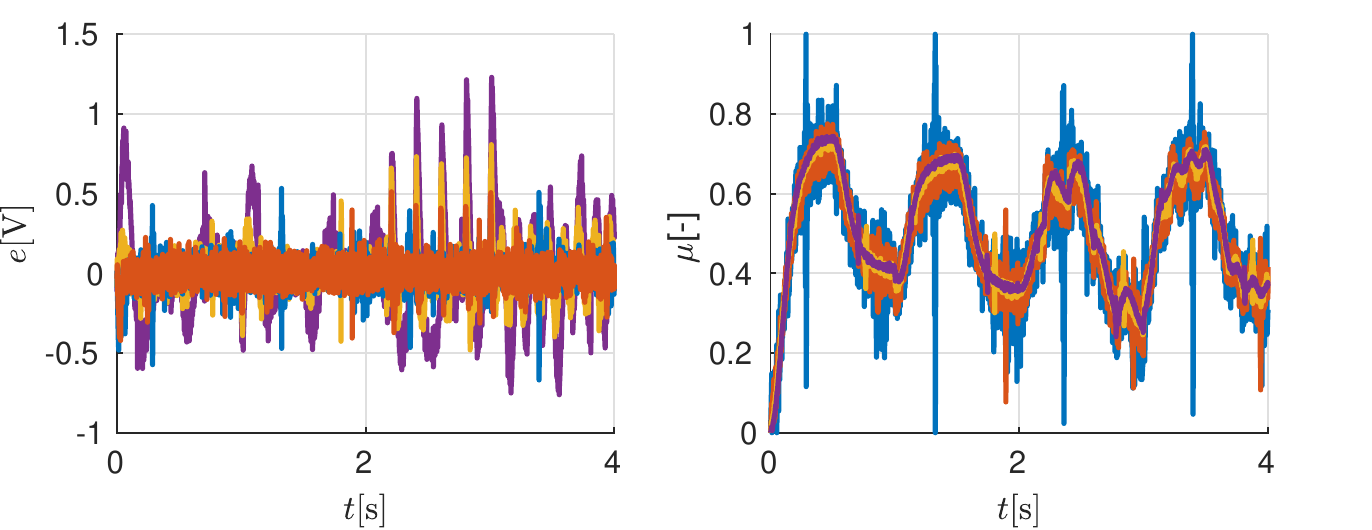}}
  \caption{Results of experiment E2c.}
  \label{figExp2c}
\end{figure}

Next, in order to provide potential CESO users with guidelines for its construction and tuning, the influence of its design parameters was investigated. To this effect, the results of E2 are seen in Fig.~\ref{figExp2a}-\ref{figExp2c}. It should be noted that the estimated total disturbance is part of the control signal (see~\eqref{eq:generalizedController}) so its influence is explicitly visible in the control signal.

\addheaders

The results of E2a are depicted in Fig.~\ref{figExp2a}. In the case of standard ESO ($p=1$), the well-known relation from high-gain observers, discussed in the Introduction, can be noticed. Namely, with the increase of observer bandwidth $\omega_{o1}$, significant noise amplification occurs in the control signal. At the same time, a slight improvement of the control error was obtained. In the case of proposed CESO ($p=2$ and $p=3$), with the increase of $\omega_{o1}$, the amplitude of the control signal increases but no visible improvement in the control accuracy can be observed. In other words, due to multiple factors like maximum sampling frequency and noise characteristics, increasing the observer bandwidth $\omega_{o1}$ will at some point no longer provide better performance. We can conclude that with the CESO one can achieve better control performance for wider range of $\omega_{o1}$ values, compared to the results of standard ESO ($p=1$) in Fig.~\ref{figExp2a}(a).

% The structure of the proposed ADRC is bulkier than the conventional but in return provides an additional and practically appealing degree of freedom in shaping the influence of measurement noise on the observer/controller part.

The results of E2b are depicted in Fig.~\ref{figExp2b}. In the case of standard ESO ($p=1$), it is clear that increasing the controller bandwidth $k$ improves the control accuracy while keeping a significant, undesired level of control signal and noise therein. In the case of proposed CESO ($p=2$ and $p=3$), increasing the controller bandwidth $k$ results in comparable control errors while retaining similar level of control signal. Due to the characteristics of CESO, it is possible to obtain better control performance for wider range of $k$ values, compared to the results obtained for the standard ESO in Fig.~\ref{figExp2b}(a).

%%%%%%%%%
%%%%%%%%%
%%%%%%%%%

%%%%%%%%%
%%%%%%%%%
%%%%%%%%%

The results of E2c are depicted in Fig.~\ref{figExp2c}. In the case of CESO ($p=2$), increasing $\alpha$ improves both the tracking accuracy and noise suppression in the control signal. However, in the case of CESO ($p=3$), increasing $\alpha$ keeps improving the noise suppression in the control signal but at some point deterioration in the tracking accuracy can be spotted. It results from a fact that, in this case, the observer bandwidth $\omega_{o1}$ is set too small, which makes the observer not providing fast-enough and accurate-enough estimate of the first state variable of the extended state vector.

\begin{figure}[t]
	\centering
	\includegraphics[width=0.485\textwidth]{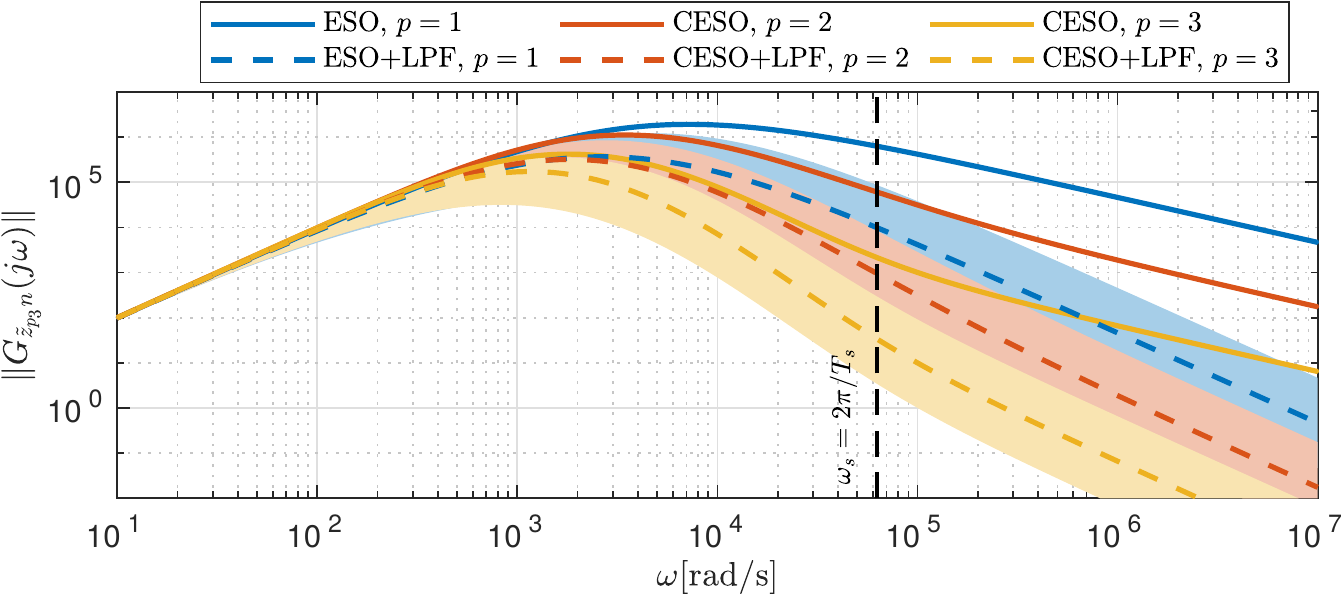}
	\caption{\color{black} Bode diagram representing the module of $G_{\extendedStateObservationErrorCascadeLevelElement{p3}\measurementNoise}(j\omega)$.}
	\label{fig:gfn}
% 	\vspace{-0.3cm}
\end{figure}

\color{black} Let us now focus on some frequency-domain insights concerning experiment E3. An algebraic transformation of \eqref{eq:generalizedController}, using \eqref{eq:observationError}, allows to write down the form of a generalized controller utilizing $p$-level cascade observer, that is directly dependent on the observation error of total disturbance $\extendedStateObservationErrorCascadeLevelElement{p3}$, i.e., $\controlSignal = \cgpEstimate^{-1}(\extendedStateElement{1} - \extendedStateObservationErrorCascadeLevelElement{p3} + \stabilizingController)$. The transformation of \eqref{eq:extendedStateAggregatedObservationErrorDerivative} into Laplace domain allows to write that for every $\cascadeLevel\geq1$% : \cascadeLevel\in\integerNumbers$
\begin{align}
    \tilde{Z}_{p3}(j\omega) = G_{\extendedStateObservationErrorCascadeLevelElement{p3}\measurementNoise}(j\omega)N(j\omega) + G_{\extendedStateObservationErrorCascadeLevelElement{p3}\extendedStateElement{3}}(j\omega)Z_{3}(j\omega),
    \label{eq:gzn}
\end{align}
where $\tilde{Z}_{p3}(j\omega)$ and $Z_{3}(j\omega)$ are the Laplace-domain equivalents of signals $\extendedStateObservationErrorCascadeLevelElement{p3}(t)$ and $\extendedStateElement{3}(t)$. Application of the LPF
\begin{align}
    G_{\textrm{LPF}}=1/(s\tau+1), \quad \tau>0,
\end{align}  \addheaders
at the output of the converter (in order to filter-out measurement noise) affects the total disturbance signal with a filtered-out parts of the measured signal $\extendedStateElement{1}$ and results in a following extended form of \eqref{eq:gzn}, i.e.,
\begin{align}
    \tilde{Z}_{p3}(j\omega) &= G_{\extendedStateObservationErrorCascadeLevelElement{p3}\measurementNoise}(j\omega)N(j\omega) + G_{\extendedStateObservationErrorCascadeLevelElement{p3}\extendedStateElement{3}}(j\omega)Z_{3}(j\omega) \nonumber \\
    &+G_{\extendedStateObservationErrorCascadeLevelElement{p3}\extendedStateElement{1}}(j\omega)Z_{1}(j\omega),
    \label{eq:gzn_lpf}
\end{align}
where $Z_{1}(j\omega)$ corresponds to signal $\extendedStateElement{1}(t)$ after Laplace transformation. According to \cite{khalil2016}, the high-gain extended observer performance should not be substantially affected for small enough values of time constant $\tau$ of the low-pass filter. We assume that $\tau$ has been chosen appropriately, and hence focus on the noise-connected characteristics of the ADRC with analyzed observers. The amplification of particular frequencies of the measurement noise using ESO and CESO ($p=2,3$) with parameters $\observerBandwidthMultiplier=3$ and $\lambda=3600$rad/s (see Table~\ref{tab:obsbandparamSPECIFIC}) has been presented in Fig.~\ref{fig:gfn}. The dashed lines represent the magnitude of $G_{\extendedStateObservationErrorCascadeLevelElement{p3}\measurementNoise}$ when a low-pass filter was applied while the regions with corresponding colors illustrate the set of characteristics that would be obtained for a practically useful set of values $\tau\in[0.0001, 0.01]$s, where the bottom edge corresponds to $\tau=0.01$s and the top edge corresponds to $\tau=0.001$s.

\begin{figure}[t]
	\centering
	\includegraphics[width=0.49\textwidth]{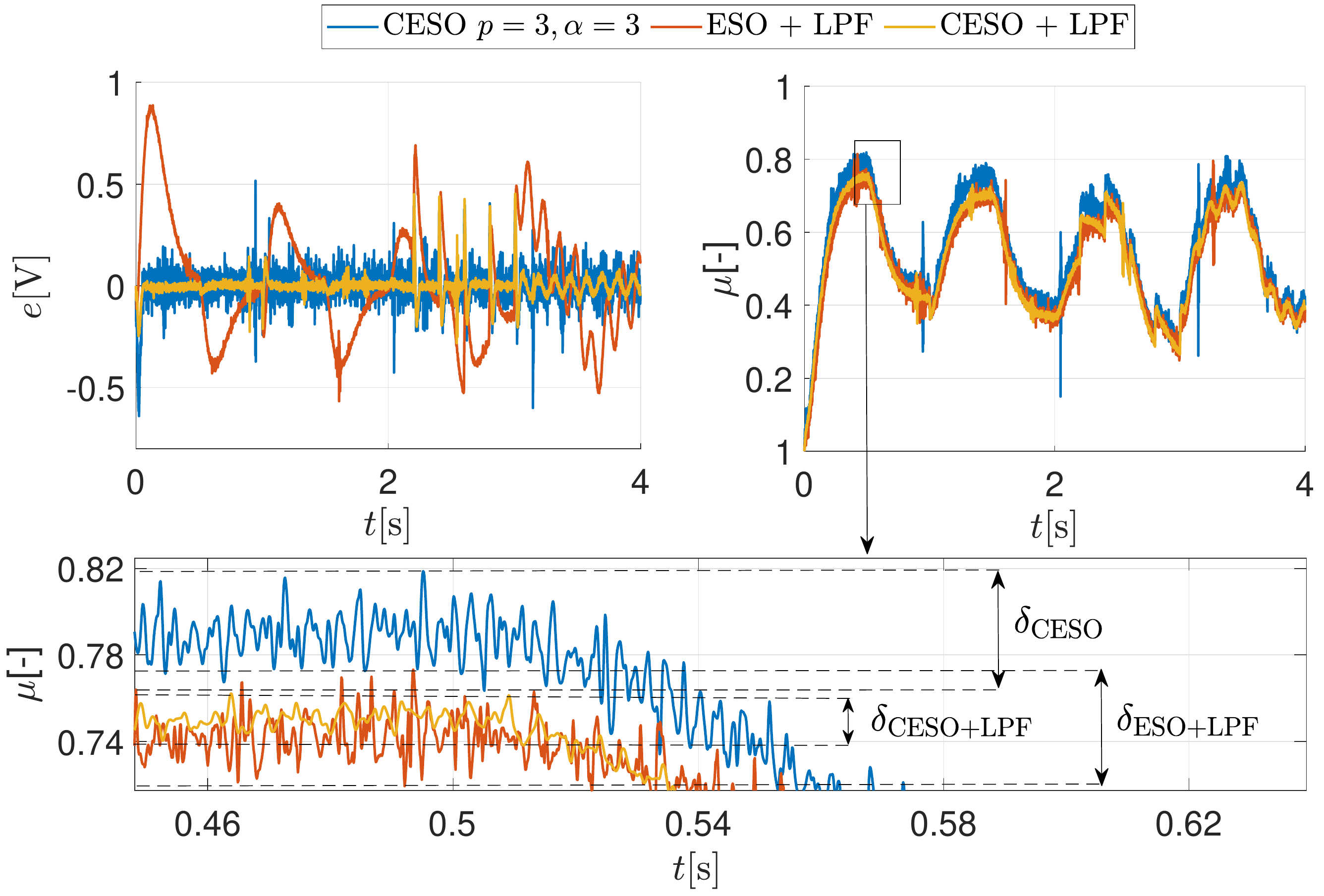}
	\caption{\textcolor{black}{Results of experiment E3.}}
	\label{fig:figExp3}
% 	\vspace{-0.3cm}
\end{figure}

% The frequency spectrum of measurement noise affecting the output of a discrete time system can include every $\omega\leq\frac{1}{2}\omega_s = \frac{\pi}{T_s}$.
Looking at Fig.~\ref{fig:gfn}, one can notice that the maximal value of $\|G_{\extendedStateObservationErrorCascadeLevelElement{p3}\measurementNoise}\|$ for CESO ($p=3$) without output filtering was similar, or smaller, compared to the characteristics obtained with the conventional ESO with LPF with $\tau=0.001$s, so the expected content of the measurement noise in signal $\extendedStateObservationErrorCascadeLevelElement{p3}$ affecting the control signal should be similar, or lower. This observation was validated by time-domain results of experiment E3, presented in Fig.~\ref{fig:figExp3}, where the amplitude of noise-dependent oscillations is $\delta_{\textrm{ESO+LPF}} \approx \delta_{\textrm{CESO}} \approx 0.05$. The presented values of the control error illustrate the essential difference in the measurement noise handling by the CESO, compared to the use of a LPF. The proposed cascade observer structure suppresses the effect of measurement noise amplification in the control signal but does not change the  noise level at the output, while the use of a LPF decreases the level of measurement noise at the output but does not change the noise amplification feature of the high-gain ESO. In order to improve the overall performance of the control system, in terms of robustness against measurement noise, a LPF can be utilized along CESO. Such example is illustrated in Fig.~\ref{fig:figExp3}, where the combination of CESO and LPF achieves the amplitude value $\delta_{\textrm{CESO+LPF}}\approx0.02$, which is smaller than the aforementioned $\delta_{\textrm{ESO+LPF}}$ and $\delta_{\textrm{CESO}}$.
\color{black}

\textcolor{black}{In order to summarize the results obtained in this work and allow for their quick assessment, Table~\ref{tab:TaxonomyTable} compares the standard ESO with the proposed CESO using selected criteria.}

\begin{table*}[th]
\centering
\caption{\textcolor{black}{Comparison between standard ESO- and proposed CESO-based control with selected criteria.}}
\begin{tabular}{>{\color{black}}c >{\color{black}}c >{\color{black}}c >{\color{black}}c}
\hline\hline
\diagbox{Criterion}{Observer type} & Standard ESO ($p=1$) & Cascade ESO ($p=2$) & Cascade ESO ($p=3$)                                \\ \hline
% Cascade level                                                    & $p=1$            & $p=2$     & $p=3$                                   \\
Observer tuning methodology                          & parameterization~\cite{GaoLESO}          & \multicolumn{2}{c}{\textcolor{black}{parameterization~\cite{GaoLESO}~$+$~Table~\ref{tab:obsbandparamSPECIFIC}}}  \\
Design parameters                      & $k,\hat{b},\omega_{o1}$    & $k,\hat{b},\omega_{o1}$, $\alpha$ &  $k,\hat{b},\omega_{o1}$, $\alpha$ \\
Total disturbance estimation quality (Fig.~\ref{fig:E1})                          & \textbf{---}          & $\nearrow$     & $\nearrow\nearrow$                       \\
Control quality $\left(\smallint|e(t)|dt~\text{in~Table~\ref{tab:qualityCriteria2}}\right)$                         & \textbf{---}          & $\nearrow$       & $\nearrow\nearrow$                    \\
Control effort $\left(\smallint|\controlSignal(t)|dt~\text{in~Table~\ref{tab:qualityCriteria2}}\right)$                         & \textbf{---}          & $\nearrow$       & $\nearrow\nearrow$                          \\
Control jittering $\left(\smallint|\dot{\controlSignal}(t)|dt~\text{in~Table~\ref{tab:qualityCriteria2}}\right)$                         & \textbf{---}          & $\searrow$       & $\searrow\searrow$                        \\
Noise content in control signal with output LPF (Fig.~\ref{fig:gfn})                       & $\searrow$ & $\searrow$ & $\searrow$\\
Implementation complexity (no. of observer state variables)                                           & $3$          & $6$       & $9$                          \\
Stability type (nominal conditions)                         & asymptotic~\cite{SenChen,lakomy2020cascade}          & \multicolumn{2}{c}{\textcolor{black}{asymptotic~(Remark~\ref{rem:nominalConditionsObservation} in Sect.~\ref{sect:MainSectCESO})}}                              \\
Stability type (non-nominal conditions)                        & practical~\cite{SenChen,lakomy2020cascade}          & \multicolumn{2}{c}{\textcolor{black}{practical~(Theorem~1 in Sect.~\ref{sect:MainSectCESO})}}                         \\
 \hline\hline
\end{tabular}
% \vspace{-0.6cm}
\label{tab:TaxonomyTable}
\end{table*}

%%%%%%%%%%%%%%%%%%%%%%%%%%%%%%%%%%%%%%%%%
%%%%%%%%%%%%%%%%%%%%%%%%%%%%%%%%%%%%%%%%%
%%%%%%%%%%%%%%%%%%%%%%%%%%%%%%%%%%%%%%%%%
\section{Conclusions}
%%%%%%%%%%%%%%%%%%%%%%%%%%%%%%%%%%%%%%%%%
%%%%%%%%%%%%%%%%%%%%%%%%%%%%%%%%%%%%%%%%%
%%%%%%%%%%%%%%%%%%%%%%%%%%%%%%%%%%%%%%%%%

%In this article, a new observer design has been proposed to enhance the control performance of an ADRC strategy controlling a DC-DC buck power converter with significant measurement noise. In particular, a novel cascade combination of extended state observers have been developed. Through a set of hardware experiments, it was shown to be capable of fast and accurate signals reconstruction, while avoiding the over-amplification of sensor noise, common in standard ADRC solutions.

An active disturbance rejection control with a novel cascade extended state observer (CESO) for DC-DC buck converters has been proposed. The validity of the new approach has been shown through a dedicated stability analysis and a set of hardware experiments. The comparison between the proposed cascade ESO-based ADRC and a standard single ESO-based ADRC showed that the former has stronger capabilities of sensor noise suppression and provides better control performance (understood as tracking accuracy and energy efficiency). The structure of the proposed ADRC is bulkier than the conventional one but in return provides an additional and practically appealing degree of freedom in shaping the influence of measurement noise on the observer/controller part.
% Additionally, simple tuning rules of the original ADRC have been preserved in the modified scheme, which simplifies its practical implementation.

% \color{black}
% limitation naszego podejscia (bardziej zlozona struktura, wiecej parametrow, choc zaproponowana intuicyjna parametryzacja)
% \color{black}

\color{black}
% Potential future work may include extending the applicability of the proposed observer to systems with time-delays~\cite{Curry1,Curry2} and/or mismatched disturbances~\cite{obs-survey2}.
\color{black}

%%%%%%%%%
%%%%%%%%%
%%%%%%%%%

\addheaders

\section{Acknowledgment}
The article was created thanks to participation in program PROM of the Polish National Agency for Academic Exchange. The program is co-financed from the European Social Fund within  the Operational Program Knowledge Education Development, non-competitive project entitled “International scholarship exchange of PhD students and academic staff” executed under the Activity 3.3 specified in the application for funding of project No. POWR.03.03.00-00-PN13 / 18. The work has been also supported by the Fundamental Research Funds for the Central Universities project no.~21620335.

%%%%%%%%%%%%%%%%%%%%%%%%%%%
%%%%%%%%%%%%%%%%%%%%%%%%%%%
%%%%%%%%%%%%%%%%%%%%%%%%%%%

% References
% \vspace*{-.25cm}
\bibliographystyle{IEEEtranTIE}
\bibliography{mybibfile}
\addheaders

\end{document}